\documentclass[11pt]{article}
\usepackage{amssymb}
\usepackage{latexsym}
\usepackage[mathscr]{eucal}
\usepackage{citesort}
\usepackage{url}
\usepackage{deleq}
\usepackage{a4}
\usepackage{cite}
\usepackage{ntheorem}
\newcommand{\myOmega}{\Omega}
\newcommand{\fourg}{{{}^4g}}
\newcommand{\nponeg}{{{}^{n+1}g}}
\newcommand{\Qpoly} {Q}
\newcommand{\DIVX} {\varphi} 
\newcommand{\pdivX}{\varphi}
\newcommand{\pdivXz}{\mathring{\varphi}}
\newcommand{\beq}{\begin{equation}}
\newcommand{\eeq}{\end{equation}}

\newcommand{\riemgz}{g_0}

\renewcommand{\hbar}{{\overline \riemgz}}

\newcommand{\const}{\mathrm{const}}

\newcommand{\zP}{{\mathring{P}}}

\newcommand{\mcJ}{{\mycal J}}
\newcommand{\mcN}{{\mycal N}}
\newcommand{\mcC}{{\mycal C}}

\newcommand{\mcW}{{\mycal W}}

\newcommand{\nablash}{D {\kern -.75 em
     \raise 1.5 true pt\hbox{{\bf/}}}\kern +.1 em}
\newcommand{\Deltash}{\Delta{\kern -.69 em
     \raise .2 true pt\hbox{{\bf/}}}\kern +.1 em}
\newcommand{\Rslash}{R{\kern -.60 em
     \raise 1.5 true pt\hbox{{\bf/}}}\kern +.1 em}

\newcommand{\Ric}{\operatorname{Ric}}

\newcommand{\hbound}{\mathring{\mathsf{H}}}
\newcommand{\cbord}{{\mathsf{C}}}

\newcommand{\maclK}{{\mathcal K}}
\newcommand{\maclKzo}{{\mathcal K}^{\bot_g}_0}

\newcommand{\hbord}{\hbound}%

\newcommand{\mcO}{{\mycal O}}

\newcommand{\mcU}{{\mycal U}}
\newcommand{\mcV}{{\mycal V}}

\newcommand{\hyp}
{{\mycal S}}

\newcommand{\mcM}{{\mycal M}}
\newcommand{\mcK}{{\mycal K}}

\newcommand{\bea}{\begin{eqnarray}}
\newcommand{\beaa}{\begin{eqnarray*}}
\newcommand{\bean}{\begin{eqnarray}\nonumber}

\newcommand{\bel}[1]{\begin{equation}\label{#1}}
\newcommand{\beal}[1]{\begin{eqnarray}\label{#1}}
\newcommand{\beadl}[1]{\begin{deqarr}\label{#1}}
\newcommand{\eeadl}[1]{\arrlabel{#1}\end{deqarr}}
\newcommand{\eeal}[1]{\label{#1}\end{eqnarray}}
\newcommand{\eead}[1]{\end{deqarr}}
\newcommand{\eea}{\end{eqnarray}}
\newcommand{\eeaa}{\end{eqnarray*}}

\newcommand{\be}{\begin{equation}}
\newcommand{\ee}{\end{equation}}

\newcommand{\tr}{\mbox{\rm tr}}

\newcommand{\eq}[1]{\eqref{#1}}
\newcommand{\Eq}[1]{Equation~(\ref{#1})}

\newcommand{\Eqs}[2]{Equations~(\ref{#1})-\eq{#2}}

\DeclareFontFamily{OT1}{rsfs}{}
\DeclareFontShape{OT1}{rsfs}{m}{n}{ <-7> rsfs5 <7-10> rsfs7 <10->
rsfs10}{} \DeclareMathAlphabet{\mycal}{OT1}{rsfs}{m}{n}

\usepackage{amsmath} 

\renewcommand{\theequation}{\thesection.\arabic{equation}}

\let\ssection=\section

\renewcommand{\section}{\setcounter{equation}{0}\ssection}

\newtheorem{defi}{\sc Definition\rm}[section]

\newtheorem{Corollary}[defi]{\sc Corollary\rm}
\newtheorem{Conjecture}[defi]{\sc Conjecture\rm}

\newtheorem{Theorem}[defi]{\sc Theorem\rm}

\newtheorem{Proposition}[defi]{\sc Proposition\rm}

\newtheorem{Lemma}[defi]{\sc Lemma\!\rm}

\theorembodyfont{\upshape}
\newtheorem{Remark}[defi]{{\sc Remark}\rm}

\newtheorem{remark}[defi]{{\sc Remark}\rm}

\newcommand{\qed}{\hfill $\Box$\bigskip}

\newcommand{\proof}{\noindent {\sc Proof:\ }}

\def \Reel{\mathbb{R}}

\def \R {\Reel}

\newcommand{\mcL}{{\mycal L}}
\def \Nat{\mathbb{N}}

\def \N {\Nat}

\newcounter{mnotecount}[section]

\newcommand{\rmnote}[1]{}

\begin{document}
\title{KIDs are non-generic}
\author{Robert Beig\thanks{ESI visiting scientist. Permanent address: Institut f\"ur Theoretische
Physik der Universit\"at Wien, Boltzmanngasse 5, A-1090 Vienna,
Austria; email
  \protect\url{robert.beig@univie.ac.at}}\\
Piotr T. Chru\'{s}ciel\thanks{ESI visiting scientist. Permanent
address: D\'epartement de Math\'ematiques, Facult\'e des Sciences,
Parc de Grandmont, F37200 Tours, France. Partially supported by a
Polish Research Committee grant 2 P03B 073 24, by the Erwin
Schr\"odinger Institute, and by a travel grant from the Vienna
City Council; email \protect\url{
piotr@gargan.math.univ-tours.fr}, URL \protect\url{
www.phys.univ-tours.fr/}$\sim$\protect\url{piotr}}{$\phantom{i}$}
\\ Richard
Schoen\thanks{ESI visiting scientist. Permanent address:
Department of Mathematics, Stanford University, Palo Alto, USA.
Partially supported by the Erwin Schr\"odinger Institute;
email \protect\url{schoen@math.stanford.edu}}\\
 \\
Erwin Schr\"odinger Institute\\ Vienna, Austria}
\date{}

\maketitle


\begin{abstract}
We prove that the space-time developments of generic solutions of
the vacuum constraint Einstein equations do not possess any global
or local Killing vectors, when Cauchy data are prescribed on an
asymptotically flat Cauchy surface, or  on a compact Cauchy
surface with mean curvature close to a constant, or for CMC
asymptotically hyperbolic initial data sets. More generally, we
show that non-existence of global symmetries implies, generically,
non-existence of local ones. As part of the argument, we prove
that generic metrics do not possess any local or global conformal
Killing vectors.

AMS 83C05, PACS 04.20.Cv
\end{abstract}


\section{Introduction}

Let $P$ be the linearisation of the general relativistic
constraints map, as defined by  \eq{1} below. Recall that a
\emph{Killing Initial Data} (KID) is a couple $(N,Y)$, defined on
a spacelike hypersurface, where $N$ is a function and $Y$ is a
vector field, such that $P^*(Y,N)=0$, see \eq{k1}-\eq{k2} below.
In vacuum space-times, with or without cosmological constant, KIDs
are in one-to-one correspondence with Killing vectors in the
associated space-time~\cite{MR54:4541,CollJMP}.

A local Killing vector field is a solution $X$ of the Killing
equations defined on an open subset of a
pseudo-Riemannian\footnote{\label{Fpsuedo} In our terminology a
Riemannian metric is also pseudo-Riemannian.} manifold $M$; local
conformal Killing vector fields and local KIDs are defined in an
analogous way.

When attempting to glue general relativistic initial data
sets~\cite{CIP} one is faced with the need of proving the
following:

\begin{Conjecture}\label{C1}
Generic general relativistic vacuum initial data sets have no
local KIDs.
\end{Conjecture}

%
%

The object of this paper is to establish  such a fact under some
supplementary conditions.
  For $\mcU\subset M$ let
$\mcK(\mcU)$ denote the set of KIDs on $\mcU$. We show, first,
that non-existence of global KIDs implies, generically,
non-existence of local ones:

\begin{Theorem}\label{T1global2} Let $\Lambda\in\R$, and consider the collection of
vacuum initial data sets with cosmological constant $\Lambda$ on
an $n$-dimensional manifold $M$ with a $C^{k,\alpha}$ topology,
$k\ge k_0(n)$, for some $k_0(n)$
($k_0(3)=6$)\footnote{\label{Fko}The function $k_0(n)$ is
obtained, in dimension $n=3$, by chasing the differentiability
thresholds throughout the proof. We do not have an explicit
estimate for $k_0(n)$ if $n>3$, or  for the function $\ell_0(n)$
appearing in Theorem~\ref{Tmain} below, because some steps of the
proof in those dimensions proceed via non-constructive arguments,
see Section~\ref{Spwc}.}, $\alpha\in (0,1)$. Let $(K_0,g_0)$ in
this collection be such that
\bel{ngkc} \mcK(M)=\{0\}\;.\ee
\begin{enumerate}
\item Let $p\in M$ and consider the  set
$$\mcV_p=\{ \mbox {vacuum initial data such that
$\mcK(\mcU)=\{0\}$ for any neighborhood $\mcU$ of $p$} \}\;.$$
Then $\mcV_p$ is open and dense in a neighborhood of $(K_0,g_0)$.
\item Define further:
$$\mcV= \{\mbox{vacuum initial data such that $\mcK(\mcU)=\{0\}$ for any open
subset $\mcU$ of $M$}\}\;.$$ Then $\mcV$  is of second category in
a neighborhood of $(K_0,g_0)$.
\end{enumerate}
Identical results hold in the class of initial data with fixed
constant $\tr_g K$, as well as in the class of time symmetric
initial data $K\equiv 0$.
\end{Theorem}

(Recall that a set is of {\em second category} if it contains a
countable intersection of open dense sets;  in complete metric or
Fr\'echet  spaces such sets are dense.)

The $C^{k,\alpha}$ topology in Theorem~\ref{T1global2}, as well as
in the remaining results below unless explicitly stated otherwise,
can be understood as follows: one chooses some smooth complete
Riemannian metric $h$ on $M$, which is then used to calculate
norms of tensors and their $h$-covariant derivatives. Other
choices are possible, and this is discussed in more detail in
Appendix~\ref{Atopo}.

 One expects that for
generic initial data the no-global-KIDs condition \eq{ngkc} of
Theorem~\ref{T1global2} will be satisfied.
 Attempts to prove that
require analytical tools which impose restrictions on the
geometry. We concentrate therefore on  three cases which seem to
us to be the most important ones from the point of view of
applications: compact manifolds without boundary, or
asymptotically flat initial data sets, or conformally
compactifiable initial data sets. Our next main result, when used
in conjunction with Theorem~\ref{T1global2}, establishes
Conjecture~\ref{C1} in those cases:

\begin{Theorem}\label{T1global} Consider the following collections of
 vacuum initial data sets: \begin{enumerate} \item $\Lambda=0$ with an
 asymptotically flat region, or \item
$\tr_g K=\Lambda=0$ with an asymptotically flat region, or \item
$K=\Lambda=0$ with an asymptotically flat region, or
\item with a conformally compactifiable region in which
$\tr_g K$ is constant, or
\item the trace of $K$ is constant and the
underlying manifold $M$ is compact, with \bel{taulamc} (\tr_g K)^2
\ge \frac
 {2n}{(n-1)}
 \Lambda\;,\ee
 or
\item $K=0$, $M$  is compact, and the curvature scalar $R$ satisfies $R=2\Lambda \le 0$,
\end{enumerate} with a
$C^{k,\alpha}\times C^{k,\alpha}$  (weighted in the non-compact
region) topology, with $k\ge k_0(n)$ for some $k_0(n)$ ($k\ge 6$
if $n=\dim M=3$). For each such collection the subset of vacuum
initial data sets without global KIDs is open and dense.
\end{Theorem}

The weights in the asymptotic region should be chosen so that the
metrics approach the Euclidean one as $r^{-\beta}$, for some
$\beta\in (0,n-2]$. In the conformally compactifiable regions a
topology as in~\cite[Theorem~6.7]{ChDelayHilbert} with $0\le
t<(n+1)/2$ should be used.

It would be of interest to have a version of points 4 and 5
without the CMC condition. Since the collection of initial data
sets which have no global KIDs is open (see
Proposition~\ref{Pclosed} below), for compact manifolds the proof
of Theorem~\ref{T1global} also provides a large open collection of
initial data sets which are close to CMC data and which have no
global KIDs. However, the general case remains open. We think that
the removal of the CMC condition in point 5 is the most notable
problem left open by our paper.

Somewhat surprisingly, the above results require a considerable
amount of non-trivial work. We first show that generic metrics
have no local conformal Killing vectors, or local Killing
vectors\footnote{The only related result known to us in the
literature is in \cite{Ebin}, where it is shown that on a compact
boundaryless manifold the set of Riemannian metrics without
nontrivial isometries is open and dense. The argument given there
does not seem to be useful to get rid of local Killing vector
fields, and makes essential use of the fact that $M$ is compact
without boundary. Moreover it is not clear how to adapt it to
account for conformal Killing vectors, or for KIDs.}. This is done
by reducing the problem to a finite system of linear algebraic
equations for the candidate vector, as well as a few of its
derivatives, at a given point. While the argument is conceptually
straightforward, there is some messy algebra involved when one
wishes to show that those algebraic equations lead to the desired
conclusion \emph{for at least one metric}. This result is then
used in the proof of Theorem~\ref{T1global}. A similar argument is
used for local KIDs, with an appropriately messier algebra. That
would have settled the problem, if not for the fact that we want
initial data satisfying the constraint equations. In order to take
care of that we first use Taylor expansions to construct
approximate solutions of the constraint equations near a point
$p$. The gluing techniques of
Corvino-Schoen~\cite{CorvinoSchoenprep} type, as extended
in~\cite{ChDelayHilbert,ChDelay}, are then used to go from an
approximate solution to a real one, establishing
Theorem~\ref{T1global2}.

 Some comments on the organisation of this paper are in order.
The heart of our analysis lies in Section~\ref{Sexact}, where we
show how to perturb solutions of the vacuum constraint equations
to solutions without KIDs, preserving the constraint equations.
This requires several preliminary results, such as a) perturbing
initial data to get rid of KIDs, without necessarily satisfying
any constraint equations, and b) perturbing metrics to get rid of
conformal Killing vectors. The argument needed for a) is presented
in Section~\ref{SnoKIDSnearp} in dimension three, and in
Section~\ref{Spwc} in all dimensions. The advantage of the
argument in Section~\ref{SnoKIDSnearp} is that it gives an
explicit differentiability threshold for the construction, in the
physically important case $n=3$, while the one in
Section~\ref{Spwc} leads to some uncontrollable, dimension
dependent threshold. In Section~\ref{Snoskids} we show how to get
rid of KIDs in the time-symmetric case, while remaining in the
time-symmetric class.
 In
Section~\ref{SmnoCKV} we construct functions that control
existence, or lack thereof, of conformal Killing vector fields in
dimension three. This result is the key for getting rid of KIDs on
CMC initial data sets; it also sets the stage for the structure of
the argument for KID-removal. As before, the higher dimensional
proof is carried out in Section~\ref{Spwc}, with some non-explicit
differentiability threshold. In Section~\ref{SmnoKV} we construct
the corresponding functions for controlling Killing vectors. Here
we obtain explicit differentiability thresholds in all dimensions.
Perturbations removing conformal Killing vectors do of course
remove Killing vectors as well, but the differentiability
thresholds we obtain in the Killing case are explicit in all
dimensions, and smaller than the corresponding conformal Killing
threshold in dimension three. In Section~\ref{Scategory} we show
how the local perturbation arguments of the previous sections can
be translated into category-type statements. This leads
immediately to the question of topologies appropriate in our
context, this is briefly discussed in Appendix~\ref{Atopo}.
Appendix~\ref{Sltg} presents a monodromy-type argument for
analytic overdetermined PDE systems, needed in the proofs of
Section~\ref{Spwc}. All the results just described join forces in
Section~\ref{Sproofs}, where Theorems~\ref{T1global2} and
\ref{T1global} are established.

\section{Metrics without conformal Killing vectors near a point, $n$=3.}
\label{SmnoCKV}  We start with some preliminaries. Unless
explicitly specified otherwise we assume in this section that
dimension equals three.  Recall that the Schouten tensor $L_{ij}$
is given by
\beq
\label{L} L_{ij} = R_{ij} - \frac{1}{4} g_{ij} R\;, \eeq where
$g_{ij}$ is a pseudo-Riemannian metric and $R_{ij}$ and $R$ are
respectively its Ricci and scalar curvature. Furthermore we define
the Cotton tensor $B_{ijk}$
\beq
\label{cotton} B_{ijk} = L_{i[j;k]}\;. \eeq The tensor $B_{ijk}$
has the following algebraic properties
\beq
\label{sym} B_{ijk} = B_{i[jk]}\;, \quad  B^i_{\phantom{i}ik} =
0\;, \quad B_{[ijk]} = 0\;, \eeq which makes five degrees of
freedom per space point. It also satisfies
\beq
\label{bianchi} B_{i[jk;l]} = 0 \;. \eeq Equivalently, we can take
\beq
\label{york} H_{ij} = \epsilon^{kl}_{\phantom{kl}i} B_{jkl}
\;.\eeq The tensor $H_{ij}$ is symmetric, tracefree and
divergence-free.

Suppose a metric has a conformal Killing vector $X$,
\beq
\label{conf} D_i X_j + D_j X_i = \frac{2}{3} \pdivX g_{ij} \;,
\eeq where $\DIVX $ here is the divergence of the vector field
$X$. Then it has to be the case that
\beq
\label{lieB} \mathcal{L}_{X} B_{ijk} = 0 \;.\eeq The reason is
that the map $Cotton$ sending a metric to its Cotton tensor
satisfies $\Phi^*Cotton[g] = Cotton[\Phi^*g]$ for any map $\Phi$
of $M$ into itself. One now applies this relation to the case
where $\Phi$ is a one- parameter family of diffeomorphisms
generated by a conformal Killing vector $X$. Taking the derivative
with respect to the parameter and using that the map $Cotton$ is
invariant under conformal rescalings of the metric one obtains
(\ref{lieB}). An equivalent form of (\ref{lieB}) is the relation
\beq
\label{lieH} \mathcal{L}_{X} H_{ij} = -\frac{1}{3}\pdivX H_{ij}
\;, \eeq Taking cyclic permutations of the equation obtained by
differentiating \eq{conf} one has
\beq
\label{cor1} D_i F_{jk} = - R_{jki}{}^{l} X_l + \frac{2}{3} \pdivX
_{[j}g_{k]i}\;, \eeq where we have defined $F_{ij} = D_{[i}
X_{j]}$ and $\pdivX_i = D_i \pdivX$. The Lie derivative of the
tensor $R_{ij}-Rg_{ij}/2(n-1)$ (here, for future reference, we
work in general dimension $n$) equals
\bel{gendimLXL}\mcL_X\left(R_{ij}-\frac{R}{2(n-1)}g_{ij}\right)= -
\frac{(n-2)}n D_iD_j\varphi \;.\ee In dimension $n=3$, to which we
return now, this reads
\beq
\label{cor2} D_i \pdivX_j = -3 \mathcal{L}_X L_{ij}\;. \eeq The
identities (\ref{cor1})-(\ref{cor2}), together with the relation
\bel{cor2.0} D_i X_j = F_{ij} + \pdivX g_{ij}/3\;,\ee imply that a
conformal Killing vector, for which the quantities $$(X^í,
F_{ij}\;, \DIVX , \pdivX_i)$$ are all zero at the point $p$, has
to vanish in a neighborhood of $p$. Using (\ref{cor1}),
(\ref{lieH}) takes the form
\beq
\label{lieH1} X^k D_k H_{ij} + 2 F{}_{(i}{}^{k} H_{j)k} + \pdivX
H_{ij} = 0\;. \eeq Next we take a derivative of (\ref{lieH1}) with
the result that
\beq
\label{DH} F{}_l{}^{k} D_k H_{ij} + \frac{4}{3}\pdivX D_l H_{ij} +
X^k D_lD_k H_{ij} + 2 (D_lF{}_{(i}{}^{k})H_{j)k} + 2F_{(i}{}^k
D_{\vert l\vert}H_{j)k} + \pdivX _l H_{ij} = 0\;. \eeq

We are ready now to pass to the proof of the main result of this
section:
\begin{Theorem}
\label{TnoCKV} Let $(M,g)$ be a smooth three dimensional
pseudo-Riemannian$^{\mbox{\scriptsize \ref{Fpsuedo}}}$ manifold.
\begin{enumerate}\item There exists a non-trivial homogeneous
polynomial $$Q(\cdot, \cdot, \cdot):\R^6\times
\R^{3\times 6}\times \R^{3\times 3\times 6}\to\R$$ 
such that if
$$Q(H, D H, D^{2} H)(p)\ne 0$$ (the $\R^6$ arises here because $H$ is symmetric), then there
exists a neighborhood $\mcO_p$ of $p$ on which there are no local
conformal Killing vectors. \item Let $\myOmega$ be a neighborhood
of $p\in M$. For any $k\ge 5$ and $\epsilon>0$ there exists a
metric $g'\in C^\infty(M)$ such that
$$\|g-g'\|_{C^{k}(\bar \myOmega)}<\epsilon\;,$$
with $g-g'$ supported in $\myOmega$, and such that $Q(H', D H',
D^{2} H')(p)$  does not vanish.
\end{enumerate}\end{Theorem}

\begin{remark}
A corresponding result in higher dimensions is proved in
Theorem~\ref{Tgen}.
\end{remark}
\begin{remark}
 Recall that a polynomial in the curvature tensor and its
derivatives is called \emph{invariant} if it is independent of the
frame used to evaluate its numerical value. Below we arbitrarily
choose some orthonormal basis of $T_pM$  to define $Q$, and it is
unlikely that the polynomial $Q$ defined in our proof will be an
invariant polynomial if the signature of the metric is Lorentzian;
moreover, it is not clear how to modify $Q$ to make it invariant
while preserving the claimed properties. Note that  one can view
$Q$ as a function on the frame bundle. In the Riemannian case we
let $\tilde Q$ be the integral of $ Q$ over those fibers with
respect to the Haar measure, then $\tilde Q$ is a non-trivial
invariant polynomial with the properties as above.

We note that the polynomial constructed below provides a
convenient tool to  capture the fact that a certain geometrically
defined matrix has rank larger than ten; the latter assertion
provides an equivalent invariant statement, regardless of
signature.
\end{remark}

\proof Before passing to the proof, some auxiliary results will be
useful., Let the superscript `` $\mathring{}$ " denote ``value at
the point $ p$'', {\em e.g.}\/, $D_k \mathring R_{ij}:= (D_k
R_{ij})(p)$. In the calculations that follow we will assume that
the metric has Riemannian signature. The remaining cases require
trivial modifications, which we leave to the reader. We start with
a Lemma:

\begin{Lemma}\label{LnoCKV1} Consider a metric such that
\beq
\label{choice1} \mathring g_{ij} = \delta_{ij}\;,\\\
\mathring R_{ij} = 0,
\\\  D_k \mathring R_{ij} = 0\;.
\eeq Furthermore let the second derivatives of the curvature be
such that
\beq
\label{choice2} D_k \mathring  H_{ij} = A x_k y_{(i}z_{j)} +
                              B y_k x_{(i}z_{j)} +
                              C z_k x_{(i}y_{j)}\;,
\eeq where $(x,y,z)$ form an orthonormal basis of $T_pM$ and the
three real numbers $A,B,C$ are all non-zero. Then the set of
algebraic equations for $$w:=(X_i,F_{ij}:=D_{[i}
X_{j]}\;,\varphi:=D^k X_k,\varphi_i:=D_i \varphi)(p)$$ obtained
from the equations
\beal{eqa0} &
[\mcL_X H_{ij} + \frac{1}{3} \pdivX H_{ij}](p) = 0\;, & \\ &
[\mcL_X D_k H_{ij}+ 2 C^m_{k(i}H_{j)m} + \frac{1}{3} D_k (\pdivX
H_{ij})](p)=0\;, 
& \label{eqa1}
\\ &[\mcL_X D_l D_k H_{ij}+
C^m_{kl}D_mH_{ij} + 2 (D_l C^m_{k(i})H_{j)m} +\phantom{xxxxxxxxxxxxxx}&\nonumber \\
&+ 2 C^m_{l(i}D_{|k|}H_{j)m} + 2 C^m_{k(i}D_{|l|}H_{j)m} +
\frac{1}{3}D_lD_k (\pdivX H_{ij})](p)=0\;,& \eeal{eqa2}with
$C^i_{jk}$ defined as \bel{Cdef} C^i_{jk} = \frac 13 \left( 2
\pdivX _{(j}\delta^i_{k)} - g_{jk} \pdivX^i\right) \ee implies
$w=0$.\end{Lemma}

\begin{Remark} \Eqs{eqa0}{eqa2} are necessarily satisfied by
every conformal Killing vector field $X$: \eq{eqa0} is equivalent
to \eq{lieH1}, while \Eqs{eqa1}{eqa2} are equivalent to the first
and second covariant derivatives of \eq{lieH1}.
\end{Remark}

\begin{Remark} It can be seen that
 $D_k \mathring H_{ij}$ in \eq{choice2} satisfies the
necessary algebraic requirements to arise from a metric ({\em
i.e.},\/ being symmetric in $(ij)$ and trace-free on all index
pairs, compare \eq{sym}-\eq{york}); this follows in any case from
Proposition~\ref{PnoCKV2} below.
\end{Remark}

\proof It immediately follows from Equations~(\ref{choice1}),
(\ref{choice2}) and (\ref{lieH1}) that $\mathring X = 0$.

Let $a$, $b$ and $c$ be defined as the following components of $F$
in the basis $(x,y,z)$: \beq \label{F} \mathring F_l{}^{k} = a(x_l
y^k - y_l x^k) + b(z_l x^k - x_l z^k) + c(y_l z^k - z_l y^k). \eeq
Evaluating (\ref{DH}) at $p$, and using $\mathring X = 0$ we find
that
\begin{eqnarray}
\label{DH1} \lefteqn{ 0 = [b(A + C)z_l - a(A + B)y_l] y_{(i}z_{j)}
+ [a(A + B)x_l - c(B + C)z_l] x_{(i}z_{j)} + {}}
\nonumber\\
&& {} + [c(B + C)y_l - b(A + C)x_l] x_{(i}y_{j)} + (aC z_l - bB
y_l) x_i x_j + (cA x_l - aC z_l) y_i y_j + {}
\nonumber\\
&& {} + (bB y_l - cA x_l) z_i z_j + \frac{4}{3} \pdivXz (A x_l
y_{(j}z_{i)} + B y_l x_{(j} z_{i)} + C z_l x_{(j} y_{i)}).
\end{eqnarray}
It follows by inspection that $a, b, c$ and $\pdivXz$ have all to
be zero. Differentiating (\ref{DH}) we find that
\begin{eqnarray}
\label{DDH} \lefteqn{ 0 = (D_m \mathring F_{lk}) D^k \mathring
H_{ij} + \frac{4}{3}\pdivXz_m D_l \mathring H_{ij} + {}} &&
\nonumber\\
&& {}+ 2 (D_l \mathring F_{(i\vert k}) D_{m \vert } \mathring
H{}_{j)}{}^k +  2 (D_m \mathring F_{(i \vert k}) D_{ l \vert }
\mathring H{}_{j)}{}^k + \pdivXz_l D_m \mathring H_{ij}\;,
\end{eqnarray}
where we have used the vanishing of $X^î$ and $D_iX^j$ at p.
 Next observe that, by
Equations~(\ref{choice1}) and (\ref{cor1}), there holds
\beq
\label{DFatp} D_i \mathring  F_{jk} = \frac{2}{3}
\pdivXz_{[j}g_{k]i}\;. \eeq We now insert (\ref{DFatp}) into
(\ref{DDH}) to find that
\begin{eqnarray}
\label{cor3} \lefteqn{ 0 = \frac{8}{3} \pdivXz_{(l}D_{m)}
\mathring H_{ij} - \frac{1}{3}g_{ml} \pdivXz_k D^k \mathring
H_{ij} + \frac{2}{3}\pdivXz_{(i}D_{\vert m \vert}\mathring H_{j)l}
+} &&
\nonumber\\
&& {}+
 \frac{2}{3}\pdivXz_{(i}D_{\vert l \vert}\mathring {H}_{j)m} -
\frac{2}{3} \pdivXz _k g_{i(l}D_{m)}\mathring H_j{}^k -
\frac{2}{3} \pdivXz_k g_{j(l}D_{m)}\mathring H_i{}^k.
\end{eqnarray}
Direct algebra using \eq{choice2} shows that $\pdivXz_i$ vanishes,
which is what had to be established. \qed

Let us show now that

\begin{Proposition}
\label{PnoCKV2}A metric satisfying (\ref{choice1})-(\ref{choice2})
exists.
\end{Proposition}

\proof We start with two elementary lemmata:

\begin{Lemma}
\label{LB1}Suppose we are given, on a star-shaped domain
$\myOmega$ in $(\R^n,\delta_{ij})$, a tensor field $B_{ijk}$
satisfying
\begin{eqnarray}
\label{e1}
 B_{ijk} & = & B_{i[jk]}\;, \\
\label{e2}
 B_{[ijk]}& = & 0 \;,\\
\label{e3}
 B_{i[jk,l]}& = & 0\;.
\end{eqnarray}
Then there exists a tensor field $L_{ij} = L_{(ij)}$ such that
\beq \label{e4} B_{ijk} = L_{i[j,k]} \;.\eeq If $B$ is a
homogeneous polynomial of order $p$, then $L$ can be chosen to be
a homogeneous polynomial of order $p+1$.
\end{Lemma}

\proof By Equations~(\ref{e1})-(\ref{e3}), there exists a tensor
field $M_{ij}$, not necessarily symmetric in $i$ and $j$,
satisfying (\ref{e4}) with $L_{ij}$ replaced by $M_{ij}$. {}From
(\ref{e2}) it follows that there exists a covector field
$\Lambda_i$ with $M_{[ij]} = \Lambda_{[i,j]}$. Set $L_{ij} =
M_{ij} - \Lambda_{i,j}$, then $L_{ij} = L_{(ij)}$ and satisfies
(\ref{e4}) thus proving Lemma~\ref{LB1}. The fact that solutions
can be chosen as polynomials follows from the explicit formula for
the primitive of a form used in  the proof of the Poincar\'e
Lemma. \qed

We will also need the following variation of a result of
Pirani~\cite{Pirani}:

\begin{Lemma} \label{LB2} Let $\myOmega$ be as in Lemma~\ref{LB1}
and on it a tensor field $R_{ijkl}$ having the symmetries of the
Riemann tensor and obeying the differential identity \beq
\label{bianchi1} R_{ij[kl,m]} = 0\;. \eeq Then there exists
$h_{ij} = h_{(ij)}$ such that \beq \label{riem} R_{ijlm} = 2
\partial_{[i}h_{j][l,m]}\;. \eeq If moreover $R_{ijkl}$ is a
homogeneous polynomial in the manifestly flat coordinates $\xi^i$
of order $q$, then $h_{ij}$ can be chosen as a homogeneous
polynomial of order $q+2$.
\end{Lemma}

\proof This is proved by inspection of the proof  in
Pirani~\cite[pp.~279-280]{Pirani}, using the fact that the proof
there consists of the repeated use of the Poincar\'e Lemma. \qed

 Returning to the proof of Proposition~\ref{PnoCKV2}, let $\xi$ be
coordinates on $\myOmega$ and define \beq \label{DB} B_{ijk} =
\frac{1}{2} \epsilon^m_{\phantom{m}jk} (\partial_n H_{im}) \xi^n,
\eeq where the constants $\partial_n H_{im}$ are given by the
right-hand-side of (\ref{choice2}). The field $B_{ijk}$ defined by
(\ref{DB}) obviously satisfies (\ref{e1}), while
\eq{e2}-(\ref{e3}) hold because \eq{choice2} is trace-free in all
indices. Now let $L_{ij}$ be the homogenous quadratic polynomial
guaranteed to exist by Lemma~\ref{LB1}. As \eq{choice2} is
symmetric in $i$ and $j$, the field  $B_{ijk}$ satisfies the
second equation in (\ref{sym}). This implies \beq \label{bianchi3}
\partial^j L_{ij} = \partial_i L,
\eeq where $L = \delta^{ij} L_{ij}$. Consider the field $S_{ijkl}$
defined by \beq \label{riem2} S_{ijkl} = 2 \delta_{k[i}L_{j]l} - 2
\delta_{l[i}L_{j]k} \;,\eeq it is a homogeneous quadratic
polynomial in $\xi$ which clearly has the symmetries of a Riemann
tensor. \Eq{bianchi3} implies that \eq{bianchi1} holds, hence all
the assumptions of Lemma~\ref{LB2} are fulfilled. Let $h_{ij}$ be
the fourth order homogeneous polynomial guaranteed to exist by
Lemma~\ref{LB2}, set \beq \label{metric} g_{ij} = \delta_{ij} +
h_{ij}\;. \eeq  Since $h$ vanishes to order three, both the
Riemann tensor and its derivatives vanish at $p$, which justifies
(\ref{choice1}). Further, the Riemann tensor $R_{ijkl}$ of $g$
coincides with $S_{ijkl}$ up to terms which give zero contribution
at $p$ in all the calculations relevant here, so that it is not
difficult to show that  $g_{ij}$ satisfies (\ref{choice2}), which
proves Proposition~\ref{PnoCKV2}.
 \qed

We can now pass to the

\medskip

\noindent {\sc Proof of Theorem~\ref{TnoCKV}:} Consider the linear
map $L$ which to $$w=(X_i\;,\;F_{ij}:=D_{[i}
X_{j]}\;,\;\varphi:=D^k X_k\;,\;\varphi_i:=D_i \varphi)(p)\in
\R^{10}$$ assigns
\beaa \R^{10}\ni w \to Lw
& := & \Big(\mcL_X H_{ij} + \frac{1}{3} \pdivX H_{ij}\;, \mcL_X
D_k H_{ij}+ 2 C^m_{k(i}H_{j)m}+ \frac{1}{3} D_k (\pdivX
H_{ij})\;, \\
&&  \mcL_X D_l D_k H_{ij}+
C^m_{kl}D_mH_{ij} + 2 (D_l C^m_{k(i})H_{j)m}
\\
&&+ 2 C^m_{l(i}D_{|k|}H_{j)m} + 2 C^m_{k(i}D_{|l|}H_{j)m} +
\frac{1}{3}D_lD_k (\pdivX H_{ij})\Big)(p)
\\ &&\phantom{xxxxxxxxxxxxxxxxxxxxxxx}\in
\R^{6}\otimes \R^{3\times 6}\otimes \R^{3\times 3\times 6}\;.\eeaa
Here the Lie derivative is calculated using the usual formula for
the Lie derivative of a tensor, and then the values  of $X$ and
its derivatives as determined by $w$ are inserted. Further, the
second derivatives of $\varphi$ are eliminated using \eq{cor2}.
 It follows
from Lemma~\ref{LnoCKV1} and Proposition~\ref{PnoCKV2} that the
set of metrics for which $L$ is injective is not empty. Standard
linear algebra implies that there exists a $10\times 10$ matrix,
say $A$, constructed by listing ten appropriately chosen rows of
$L$, which has non-vanishing determinant when $H$ arises from the
metric of Proposition~\ref{PnoCKV2}. Let $Q$ be the sum of squares
of determinants of all  ten-by-ten submatrices of $L$, then $Q\ge
(\det A)^2$ and therefore $Q$ is not identically vanishing by
construction. Clearly $L$ is injective whenever $ Q$ is non-zero,
which proves point 1.

To prove point 2, let $g$ be an arbitrary metric, if $Q(p)$,
evaluated for the metric $g$, does not vanish, then the result is
true with $g'=g$. Otherwise, define \bean \mcJ_5&:=&\{\mbox{the
set of fifth jets of $g$ in normal coordinates at $p$ } \\&&
\phantom{\{} \mbox{ as $g$ varies in the set of all Riemannian
metrics}\}\;.\quad \eeal{defRtwo} This a linear space, an explicit
parameterisation of which can be found in~\cite{Thomas}. Let
$e_i$, $i=1,\ldots,N$, be any basis of
 $\mcJ_5$, thus every $j\in \mcJ_5$ can be written
as
$$j=\jmath{}^i e_i\;,$$ for some numbers $\jmath{}^i\in \R$. By definition
of $\mcJ_5$, for every $(\jmath{}^i) \in \R^N$ there exists some
Riemannian metric for which $j=\jmath{}^i e_i$. Clearly the map
$g\to (\jmath{}^i)$ is continuous in a $C^\ell (\bar \myOmega)$,
$\ell \ge 5$, topology on the set of metrics, and a small
variation of $\jmath{}^i$ can be realised by a small variation of
$g$. In a frame such that $g_{ij}(p)=\delta_{ij}$, the map that
assigns to the fifth jets of $g$, at $p$, the values of the
tensors $H$, $DH$, and $DD^2H$ at $p$, is a polynomial on
$\mcJ_5$.

We want to show that a small variation of $g$ will make $Q$
non-zero. Now, $Q$ is a polynomial in the $\jmath{}^i$'s. Let
$\jmath{}^i_0$ be the values of the $\jmath{}^i$'s corresponding
to the metric $g$, and suppose that we have
$$\forall \ i_1,\ldots, i_n\qquad \frac{\partial^{i_1+\ldots+i_N} Q}{\partial^{i_1} \jmath{}^1\ldots
\partial^{i_N} \jmath{}^N}(\jmath{}^i_0)=0\;.$$
Then the polynomial $Q$ would identically vanish, contradicting
its construction. Hence there exists at least one of the above
partial derivatives which does not vanish, and therefore an
appropriate, no matter how small, variation of  $g$  will lead to
a non-vanishing value of $Q$ at $p$. As the argument depends only
upon the jets of $g$ at $p$, the variation can be made supported
in a ball containing $p$ with radius  as small as desired. \qed

\section{Metrics without Killing vectors near a point}
\label{SmnoKV}

Results on non-existence of Killing vectors follow of course
immediately from those on non-existence of conformal Killing
vectors, as established above. However, for Killing vectors in
dimension three the differentiability threshold of
Theorem~\ref{TnoCKV} can be lowered to three. Further, for Killing
vectors a simple proof can be given in all dimensions:

\begin{Theorem}
\label{TnoKV} Let $(M,g)$ be a $n$-dimensional pseudo-Riemannian
manifold. \begin{enumerate} \item There exists a non-trivial
homogeneous invariant polynomial $P_n[g]:=(DR, \ldots, D ^{2n+1}
R)$ of degree $n$, where $R$ is the Ricci scalar, such that if
$$P_n(DR, \ldots, D ^{2n+1} R)(p)\ne 0$$ at a point $p\in M$, then
there exists a neighborhood $\mcO_p$ of $p$ such that there are no
non-trivial Killing vectors on any open subset of $\mcO_p$. In
dimension $n=3$ there exists such a polynomial $\hat P_3$ which
depends upon $\Ric$ and $D \Ric$.
\item Let $\myOmega$ be a neighborhood of
$p\in M$. For any $k\ge 2n+1$ and $\epsilon>0$ there exists a
metric $g'$ such that
\bel{nearg}\|g-g'\|_{C^{k}(\bar \myOmega)}<\epsilon\;,\ee with
$g-g'$ supported in $\myOmega$, and such that $P_n(DR', \ldots, D
^{2n+1} R')(p)$  does not vanish. In dimension three we can
arrange for the non-vanishing of $\hat P_3(\Ric',D \Ric')(p)$
using a perturbation supported in $\myOmega$ and satisfying
\eq{nearg} for each arbitrarily chosen
 $k\ge 3$.
\end{enumerate}
\end{Theorem}

\begin{Remark}
The differentiability required above in dimension $n$ is certainly
not optimal, but it allows the simple proof below.
\end{Remark}

\begin{Remark}
The polynomial $P_n$ obtained here is completely useless from the
point of view of Killing vectors in vacuum space-times, where the
Ricci scalar vanishes. In this context it is of interest to have a
statement as above with a polynomial depending only upon the Weyl
tensor, and we prove existence of such polynomials in
Theorem~\ref{Tgen} below. Further, in Section~\ref{Sexact} we will
 construct small perturbations
of initial data which preserve the vacuum constraints.
\end{Remark}

\proof If $X$ is a Killing vector we have $\mcL_X (\Delta^k R)$=0
for all $k$, where $\Delta^k$ denotes the $k-th$ power of the
Laplace operator $\Delta$. At $p$ this gives the linear system of
equations
$$A_{ij}X^i(p)=0\;, \quad A_{ij}= D_i(\Delta^j R)(p)\;,\qquad
j=0,\ldots,n-1 \;.$$ Let $P_n = \det (A_{ij})$. If $P_n(p)$ does
not vanish, then $X(q)=0$ for all $q$ in the neighborhood of $p$
defined as $\{q:P_n(q)\ne 0\}$, hence $X\equiv 0$. It is not too
difficult to check, using Taylor expansions of the metric (point 2
of Proposition~\ref{PapP} below is useful  here),  that there
exist metrics for which $P_n\ne 0$, and  the result follows by a
repetition of the arguments of the proof of Theorem~\ref{TnoCKV}.

In  dimension $3$ the number of the derivatives of the metric
needed can be improved as follows: Let $G_{ij}=R_{ij}
-g^{kl}R_{kl}g_{ij}/2$, in the notation of Section~\ref{SmnoCKV}
we assume that
\begin{equation}
\label{einstein} \mathring G_{ij} = \lambda_1 x_i x_j + \lambda_2
y_i y_j + \lambda_3 z_i z_j\;,
\end{equation}
with $(\lambda_1 - \lambda_2)(\lambda_2 - \lambda_3) (\lambda_3 -
\lambda_1) \neq 0$. We set, as in (\ref{F}),
\begin{equation}
\label{F1} \mathring F_{ij} = 2(a x_{[i}y_{j]} + b z_{[i}x_{j]} +
c y_{[i}z_{j]})\;,
\end{equation}
so that
\begin{equation}
\label{FG} \mathring F_{(i}{}^k \mathring G_{j)k} = b(\lambda_1 -
\lambda_3) x_{(i}z_{j)} + c(\lambda_3 - \lambda_2)y_{(i}z_{j)} +
a(\lambda_2 - \lambda_1)x_{(i}y_{j)}\;,
\end{equation}
which has zero components on the diagonal. Finally we assume that
\begin{equation}
\label{DG} \mathring G_{ij;k} = \mu_1 (x_{(i}\delta_{j)k} - 2 x_k
\delta_{ij}) + \mu_2 (y_{(i} \delta_{j)k} - 2 y_k \delta_{ij}) +
\mu_3 (z_{(i} \delta_{j)k} - 2 z_k \delta_{ij})\;,
\end{equation}
where $\mu_1 \mu_2 \mu_3 \neq 0$. We now set
\begin{equation}
\label{Xatp} \mathring X = \alpha x_i + \beta y_i + \gamma z_i.
\end{equation}
Writing \eq{DG} in the form $\mathring G^1_{ij;k} + \mathring
G^2_{ij;k} + \mathring G^3_{ij;k}$, we find
\begin{eqnarray}
\label{DXG} \mathring G^1_{ij;k} \mathring X^k &=& \ \mu_1 \alpha
(-2y_iy_j-2z_iz_j-x_ix_j) +
\textrm{off diagonal terms}\;,\\
\mathring G^2_{ij;k} \mathring X^k &=& \ \mu_2 \beta
(-2x_ix_j-2z_iz_j-y_iy_j) +
\textrm{off diagonal terms}\;,\\
\mathring G^3_{ij;k} \mathring X^k &=&  \ \mu_3 \gamma
(-2x_ix_j-2y_iy_j-z_iz_j) + \textrm{off diagonal terms}\;.
\end{eqnarray}
We first consider the relation $\mcL_X G_{ij} = 0$ with $i=j$.
Then \eq{FG} gives no contribution, while from \eq{DXG} we obtain
a linear homogenous system for $(\alpha,\beta,\gamma)$ with
coefficient matrix $\Delta$ given by
\begin{displaymath}
\label{det} {\Delta} = \left( \begin{array}{ccc}
\mu_1 & 2\mu_2 & 2\mu_3 \\
2\mu_1 & \mu_2 & 2\mu_3 \\
2\mu_1 & 2\mu_2 & \mu_3
\end{array} \right)\;.
\end{displaymath}
There holds $\det(\Delta)= 5 \mu_1\mu_2\mu_3 \neq 0$. Thus, the
equation $\mcL_X G_{ij} =0$, satisfied by any Killing vector,
leads to $\alpha =
\beta = \gamma =0$. The off-diagonal components of $\mcL_X  G_{ij}
= 0$ imply now, by \eq{FG}, that $a=b=c=0$. Since \eq{einstein} is
symmetric, and \eq{DG} satisfies the linearised Bianchi
identities, the results in~\cite{Thomas} show that there exists a
metric $g_{ij}=\delta_{ij}+h_{ij}$, with $h_{ij}= O(\xi^2)$,
satisfying (\ref{einstein}) and (\ref{DG}). The proof is completed
by the same argument as already given for general $n$.
\phantom{XX}
 \qed

\section{Generic non-existence of local Killing, or conformal Killing, vector fields}\label{Scategory}

In this section we only consider three dimensional manifolds, the
reader will easily formulate an equivalent statement and proof for
local Killing vector fields in any dimension using
Theorem~\ref{TnoKV}, or for local conformal Killing vector fields
using Theorem~\ref{Tgen} below.
\begin{Theorem}
\label{Tsc} Let $M$ be a three dimensional manifold. Then
\begin{enumerate}
\item The set of pseudo-Riemannian metrics on $M$ which have no local Killing vector fields
is of second category in the $C^3$ topology.
 \item The set of pseudo-Riemannian
metrics on $M$ which have no local conformal Killing vector fields
is of second category in the $C^5$ topology.
\end{enumerate}
\end{Theorem}


\proof  We start with the following:

\begin{Proposition} \label{Pclosed}
Let $\myOmega$ be a domain in $M$. Then:
\begin{enumerate} \item The set of metrics on $\myOmega$ which have
no Killing vectors on $\myOmega$ is open in a $C^k(\bar \myOmega)$
topology, $k\ge 2$.
\item The set of metrics on
$\myOmega$ which have no conformal Killing vectors on $\myOmega$
is open in a $C^k(\bar \myOmega)$ topology, $k\ge 3$.\item The set
of initial data $(g,K)$ on $\myOmega$ which have no non-trivial
KIDs on $\myOmega$ is open in a $C^{k+1}(\bar \myOmega)\oplus
C^k(\bar \myOmega) $ topology, $k\ge 1$.
\end{enumerate}
\end{Proposition}

\begin{Remark}
\label{Rtopoclosed}  The openness established here holds for any
metrisable topology $\mycal T_k$ such that convergence in $\mycal
T_k$ implies uniform convergence in $C^k$ norm on compact sets,
with $k \ge2$ for Killing vectors, \emph{etc}; see also
Appendix~\ref{Atopo}.
\end{Remark}

\proof We will show that  \emph{existence} of Killing vectors, or
conformal Killing vectors, or KIDs, is a closed property. We start
with the slightly simpler case of conditionally compact
$\myOmega$:
\begin{Lemma}
\label{Lclosed} Proposition~\ref{Pclosed} holds if $\myOmega$ has
compact closure.
\end{Lemma}

 \proof 1. Let
$\gamma_i$ be a sequence of metrics with non-zero Killing vectors
$X(i)$. Rescaling $X(i)$ we can assume that
\bel{Xcond}\sup_{p\in\myOmega}\gamma_i(X(i),X(i))=1\;.\ee We note that Killing vectors extend by continuity
to $\overline\myOmega$, we shall use the same symbol to denote
that extension. Let $p_i\ \in \overline\myOmega$ be such that the
sup is attained, passing to a subsequence if necessary there
exists $p_*$ in $\overline \myOmega$ such that $p_i\to p_*$. Now,
Killing vectors satisfy the system of equations
\bel{XKV} D_iD_jX_k=R^\ell{}_{ijk}X_\ell\;,\ee which shows that second
covariant derivatives of all the $X(i)$'s are uniformly bounded on
$\overline \myOmega$. Interpolation~\cite[Appendix]{Hormander}
shows that the sequence $X(i)$ is uniformly bounded in $C^2$. The
existence of a subsequence converging in $C^1$ to a non-trivial
Killing vector field follows from the Arzela-Ascoli theorem.

2. The argument is essentially identical, with the following
modifications: we replace the normalisation \eq{Xcond} by
\bel{Xcond1}\sup_{p\in\myOmega}\left(|X(i)|_{\gamma_i}+|DX(i)|_{\gamma_i}\right)=1\;.\ee
\Eq{XKV} is replaced by the set of equations
\eq{cor1}-\eq{cor2.0}. Those equations easily imply boundedness of
the sequence $X(i)$ in $C^3$, leading to a converging subsequence
in $C^2$.

3. Let $(\gamma_i,K_i)$ be a sequence of metrics with non-zero
KIDs $(Y(i),N(i))$. We use the normalisation
\bel{Xcond2}\sup_{p\in\myOmega}\left(|Y(i)|_{\gamma_i}+|D
Y(i)|_{\gamma_i}+|N(i)|+|DN(i)|_{\gamma_i}\right)=1\;.\ee {}From
\eq{k2} and \eq{k1b} one obtains a uniform $C^2$ bound on
$(Y(i),N(i))$, and one concludes as before.
 \qed

 {Returning to the proof of point 1 of Proposition~\ref{Pclosed}}, let
 $\myOmega_j$ be an increasing  sequence of conditionally compact domains such
 that $\myOmega=\cup\myOmega_j$. By Lemma~\ref{Lclosed} we have
 $\mcK(\myOmega_j)\ne \{0\}$ for all $j$. The restriction map
 induces an injection $i_{i,j}:\mcK(\myOmega_i)\to
 \mcK(\myOmega_j)$, $i\ge j$, so that $1\le \dim
 i_{i,1}(\mcK(\myOmega_i))$ for all $i$, with
 $ i_{i+1,1}(\mcK(\myOmega_{i+1}))\subset i_{i,1}(\mcK(\myOmega_i))\subset \mcK(\myOmega_1)$.
 It follows that
 $F:=\cap _i i_{i,1}(\mcK(\myOmega_i))\ne \{0\}$, and every
 element of $F$ extends to a globally defined Killing vector field
 on $\myOmega$.
 \qed

{\noindent \sc Proof of Theorem~\ref{Tsc}:} Let $p_i$, $i\in \N$
be a dense collection of points and let $B(p_i,1/j)$, $j\ge N_i$,
be a collection of coordinate balls with compact closure. Let
$\mcV_{i,j}$ be the set of metrics such that
$\mcK(B(p_i,1/j))=\{0\}$. By Proposition~\ref{Pclosed} the set
$\mcV_{i,j}$  is open, and it  is dense by Theorem~\ref{TnoKV}.
Then any metric in $\cap_{i,j}\mcV_{i,j}$ has no local Killing
vectors. The argument for conformal Killing vector fields is
identical, based on Theorem~\ref{TnoCKV}. \qed

\section{Three dimensional initial data sets without KIDs near a point}
\label{SnoKIDSnearp}

 We now pass to the
construction of initial-data sets without KIDS.  Let
$\mcC(K_{ij},g_{ij}):=(J_i,\rho)$ be the constraints map, \beq
\label{k6} \rho:=R + K^2 - K_{ij} K^{ij} -2\Lambda \;,\eeq \beq
\label{k7} J_i:=-2D^j(K_{ij} - K g_{ij})\;, \eeq where $\Lambda\in
\R$ is the cosmological constant. In this section, and \emph{only}
is this section, the symbol $K$ denotes the trace of $K_{ij}$; $K$
stands for the full extrinsic curvature tensor elsewhere in this
paper. Let $P$ denote the linearisation of $\mcC$, and let $P^*$
be the formal adjoint of $P$. By definition, a KID $(N, X^i)$ is a
solution of the set of equations $P^*(N,X)=0$; explicitly, in
dimension $n$ (\emph{cf., e.g.}~\cite{ChDelay}),
\beq \label{k1} D_{(i}X_{j)} = - N K_{ij} \;,\eeq  \beq \label{k2}
D_i D_j N = N (R_{ij} + K K_{ij} - 2K_{il}K_j{}^{l}) - \mcL_X
K_{ij} + \frac 1 {(n-1)}\left(\frac {J_lX^l }2  -\left(  \rho
+2\Lambda\right)N \right)g_{ij}\;.\eeq One checks that any KID
$(N,X^i)$ for which $X^i, F_{ij} = D_{[i}X_{j]}\;, N$ and
$N_i:=D_iN $ all vanish at $p$ has to be zero in a neighborhood of
$p$. This is proved in the usual way from \eq{k2} together with
\beq \label{k1b} D_{\ell}D_jX_{i} = - R_{ji\ell}{^k}X_k -
D_\ell(NK_{ij}) - D_j(NK_{\ell i} ) + D_i(NK_{j\ell})\;. \eeq
(\Eq{k1b} is obtained by considering cyclic permutations of first
derivatives of \eq{k1}.) Since $\mcL_X \Ric( g) = \Ric' (\mcL_X
g)$, the usual formula for $\Ric'$ leads to \bean \label{k3a}
\mcL_X  R_{ij} &=& \Delta (N K_{ij}) + D_i D_j(NK) - 2
D_{(i}D^l(NK_{j)l}) - 2 N R_{lijm} K^{lm} - 2 N
R_{(i}^{\phantom{(i}l} K_{j)l} \\ &=& \Delta (N K_{ij}) + D_i
D_j(NK) - 2 D^lD_{(i}(NK_{j)l})  \;.\eeal{k3} From now on we
assume that $n=3$. By taking the curl of \eq{k2} one also finds
\begin{eqnarray}
\label{k4} \lefteqn{ R{}_{lij}{}^{k} D_k N = - 2 \mcL_X
D_{[l}K_{i]j} - 2C_{j[l}^m K_{i]m}} &&
\nonumber\\
&& + 2D_{[l}\left[N(R_{i]j} + K_{i]j}K - 2 K_{i]m}K_j{}^{m})  +
\left(\frac {J_mX^m }4 -\left( \frac \rho 2 +\Lambda\right)
N\right)g_{i]j}\right] ,\phantom{XXX}
\end{eqnarray}
where  \beq \label{k5}  C_{jk}^i = -g^{in}[D_j(NK_{kn}) +
D_k(NK_{jn}) - D_n(NK_{jk})] \;.\eeq  We choose some $\alpha,
\beta , \lambda_i,
a_i\in \R$ and we consider initial data with the following
properties at $p$:  \beq \label{k8} \mathring R_{ij} =
\frac{\beta}{3}\delta_{ij}\;, \ \mathring K_{ij} =
\frac{\alpha}{3}\delta_{ij}\;, \ D_i \mathring K_{jl} = 0
\;,\eeq \beq \label{k9} D_i \mathring R_{jk} = \lambda_x x_i
y_{(j}z_{k)} + \mbox{\rm (cyclic)} , \eeq where $\mbox{\rm
(cyclic)}$ means cyclic permutations of $(x,y,z)$, and $\lambda_x
\lambda_y \lambda_z \neq 0$. (This ansatz is general enough to
lead to the required result, and simple enough so that the
calculations are manageable. We will show shortly that such
initial data exist.)
We also assume that
\beq \label{k10} D_i D_j \mathring K_{lm} = a_x x_i x_j
y_{(l}z_{m)} + \mbox{\rm (cyclic)} \;, \eeq with \beq \label{in1}
\lambda_x - \lambda_y \neq \frac{a_x^2}{\lambda_x} -
\frac{a_y^2}{\lambda_y} \eeq and
\beq \label{in2} \lambda_x + \lambda_y \neq 0\;,\ \lambda_x +
\lambda_z \neq 0\;,\   \lambda_y + \lambda_z \neq 0\;. \eeq For
further reference we note that, in local coordinates $\xi$  such
that $p$ corresponds to $\xi=0$, \eq{k8}-\eq{k10} imply
\bel{consviol1}R + K^2 - K_{ij} K^{ij} = \beta + \frac
{2\alpha^2}3 +O(\xi^2) \;,\quad D^j(K_{ij} - K g_{ij}) =
O(\xi^2)\;.\ee In particular, if 
$\beta=2\Lambda-2\alpha^2/3$ then \bel{consviol2}\rho=R + K^2 -
K_{ij} K^{ij} -2\Lambda= O(\xi^2) \;,\quad J_i=-2D^j(K_{ij} - K
g_{ij}) = O(\xi^2)\;. \ee Inserting (\ref{k8}) into (\ref{k1}) and
(\ref{k2}) we find that \bea \label{l1.a} D_{(i}\mathring X_{j)}&
= &
-\frac{\alpha}{3}\mathring N \delta_{ij}\;, \\
D_iD_j\mathring N &=& 
\left(\beta + {\alpha^2} - \frac {3\rho} 2 -3\Lambda\right)
\frac{\mathring N}3 
\delta_{ij} 
\nonumber
\\ &=& - 
\frac{\mathring N \beta }6 
\delta_{ij}
\;, \\
\Delta \mathring N &=&
- 
\frac{\mathring N \beta }2 \;. \eeal{l1.c}
Evaluating (\ref{k3}) at $p$, it
follows that
\beq
\label{k11} \mathring X_m D^m \mathring R_{ij} = \mathring N
\Delta \mathring K_{ij}
\;,\eeq and, from (\ref{k4}), that
\beq
\label{k12} \mathring N D_{[l} \mathring R_{i]j} = \mathring X_m
D^m D_{[l} \mathring K_{i]j}
 \;.\eeq
{}From (\ref{k11}) we find, using the expansion $\mathring X_i =
\alpha_x x^i + \alpha_y y^i + \alpha_z z^i$, that
\beq
\label{k13} \alpha_x \lambda_x = \mathring N a_x\;,\  \alpha_y
\lambda_y = \mathring N a_y\;,\ \alpha_z \lambda_z = \mathring N
a_z \;,\ \eeq and from (\ref{k12})
\beq
\label{k14} \mathring  N (\lambda_x - \lambda_y) = \alpha_x a_x -
\alpha_y a_y\;,\ \mathring  N (\lambda_y - \lambda_z) = \alpha_y
a_y - \alpha_z a_z\;,\ \mathring  N (\lambda_z - \lambda_x) =
\alpha_z a_z - \alpha_x a_x\;. \eeq
 Combining (\ref{k14}) with
(\ref{k13}) and using \eq{in1}, it follows that
\beq \label{k15} \mathring  N = 0 = \alpha_x = \alpha_y =
\alpha_z\;. \eeq Using \eq{k15} in the first derivative of \eq{k2}
and in \eq{k1b}, we infer that
\beq
\label{l2} D_lD_{(i}\mathring X_{j)} = -
\frac{\alpha}{3}\delta_{ij} D_l\mathring N \;, \
D_lD_iD_j\mathring N = - 
\frac \beta 6
\delta_{ij} D_l\mathring N
\;.\eeq We now take a derivative of (\ref{k3}) to obtain (recall
that $F_{ij}$ is the anti-symmetric part of $D_iX_j$)
\beq
\label{k16} \mathring  F_{km} D^m \mathring  R_{ij} + 2 \mathring
F_{(i\vert}{^m}D_{k\vert} \mathring R_{j)m} = (D_k \mathring N)
\Delta \mathring  K_{ij} + 2 (D_l \mathring  N) D_k D^l \mathring
K_{ij} - 2 (D_l \mathring  N) D_k D_{(i} \mathring K_{j)}{}^{l}\;.
\eeq 
Somewhat surprisingly, all terms
involving $\alpha$  and $\beta$ have dropped out. We have to
compute the different terms entering (\ref{k16}). Writing
$\mathring F_{ij}$ as
\beq \label{k17} \mathring  F_{ij} = A_x y_{[i} z_{j]} + \mbox{\rm
(cyclic)}\; , \eeq we obtain \beq \label{k18} \mathring  F_{km}
D^m \mathring  R_{ij} = \frac{1}{2} \lambda_x (A_y z_k - A_z y_k)
y_{(i} z_{j)} + \mbox{\rm (cyclic)} \; , \eeq \beq \label{k19}
\mathring  F_{im} D_k \mathring R_j{}^m = \frac{1}{4}(\lambda_x
x_k y_j + \lambda_y y_k x_j)(A_x y_i - A_y x_i) + \mbox{\rm
(cyclic)}\;  . \eeq Also, decomposing $D_i\mathring N = u_x x_i +
u_y y_i + u_z z_i$, we have that \beq \label{k20} (D_k \mathring
N) \Delta \mathring  K_{ij} = (u_x x_k +  u_y y_k + u_z z_k) (a_x
y_{(i} z_{j)} + \mbox{\rm (cyclic)} )\; . \eeq and \beq
\label{k21}  2 (D_l \mathring  N) (D_k D^l  \mathring K_{ij} - D_k
D_{(i} \mathring  K_{j)}{}^{l})= u_x (2 a_x x_k - a_y y_k - a_z
z_k) y_{(i} z_{j)} + \mbox{\rm (cyclic)} \;. \eeq We now insert
Equations~(\ref{k18})-(\ref{k21}) into (\ref{k16}). Contracting
the resulting equation first with $x^k y^i z^j$ and cyclic
permutations thereof, one sees that $u_x, u_y, u_z$ have to
vanish. Contracting, then, with terms of the form $x^k x^i y^j,
x^k x^i z^j, y^k y^i x^j$, \emph{etc.}, we see that $A_x, A_y,
A_z$ are also zero, due to (\ref{in2}). Thus $(N, X^i)$ is zero
near $p$. We have thus proved:

\begin{Lemma}\label{L3.1} Consider an initial data set
$(g_{ij}\;,K_{ij})$ satisfying Eqs. (\ref{k8})-(\ref{k10})
together with the conditions on the coefficients spelled out
above. For any $\alpha,\beta,\Lambda\in\R$ the algebraic equations
for $r=(X_i,F_{ij},N,D_iN)$ obtained from (\ref{k1})-(\ref{k2}) by
taking derivatives up to order two imply the vanishing of $r(p)$.
\qed
\end{Lemma}

 We also have the following KID-analogue of
Proposition~\ref{PnoCKV2}:

\begin{Proposition}
\label{PnoKIS} 1. A pair $(g_{ij},K_{ij})$ satisfying
(\ref{k8})-(\ref{k10}) exists.\\ 2. Further, one can choose
$g_{ij}=\delta_{ij}+ h_{ij}$ and $K_{ij}$ so that, in local
coordinates $\xi$, the tensor fields $g_{ij}$ and $K_{ij}$ satisfy
the vacuum constraints up to terms which are of $O(|\xi|^2)$.
\end{Proposition}

\proof  By Lemma \ref{LB2} we can find $h_{ij}$ of order
$O(|\xi|^2)$, so that (\ref{k8})-(\ref{k9}) are satisfied. For
$K_{ij}$ we choose \beq \label{K} K_{ij} = \frac \alpha 3
\delta_{ij}+\frac{1}{2} (\mathring K_{ijlm} + \frac{\alpha}{3}
\partial_l \partial_m \mathring h_{ij}) \xi^l \xi^m, \eeq where
the second term on the right-hand side of (\ref{K}) is given by
the right-hand side of (\ref{k10}). One checks that \eq{k10} is
valid. Point 2. follows from \eq{consviol2}. \qed

We are ready now to prove:

\begin{Theorem} \label{TnoKIDs} Let $\alpha,\beta\in\R$, $p\in M$,
and consider the collection of all three dimensional data sets
$(M,K_{ij},g_{ij})$ with $(K_{ij},g_{ij})\in C^k\times C^{k+1}$,
$k\ge 3$, with the trace $K(p)$ of $K_{ij}(p)$ equal to $\alpha$,
and with $R(p)=\beta$.
\begin{enumerate}
\item There exists a non-trivial homogeneous invariant polynomial $\Qpoly
 [K_{ij},g_{ij}]:=\Qpoly (R_{ij},DR_{ij},D^2R_{ij},K_{ij},D K_{ij}, D^2 K_{ij},D^3 K_{ij})$
 such that if $$\Qpoly [K_{ij},g_{ij}](p)\ne 0$$ at a
point $p\in M$, then there exists a neighborhood $\mcO_p$ of $p$
for which there exist no non-trivial KIDS on any open subset of
$\mcO_p$. \item Let $\myOmega$ be a  domain in $M$ with $p\in
\myOmega$. a) There exists a variation $(\delta K_{ij}, \delta
g_{ij})\in {(C^{\infty}\times C^{\infty})(\myOmega)}$, compactly
supported in $\myOmega$, such that  $\Qpoly [K_{ij}+
\epsilon\delta K_{ij}, g_{ij}+\epsilon\delta g_{ij}](p)\ne 0$ for
all $\epsilon $ small enough. b) The variation can be chosen so
that it preserves the value of $R(p)$ and of $K(p)$. One can
further arrange for the trace of $K_{ij}+ \epsilon\delta K_{ij}$
to be equal to $K$ throughout $\myOmega$ when $K$ is a constant.
\item If $(K_{ij},g_{ij})$ is vacuum (with perhaps non-zero cosmological
constant) with $(K_{ij},g_{ij})\in C^{k+\ell+1}\times
C^{k+\ell+2}$,
$\ell\ge 0$, then for any $p\in \myOmega$ 
the variation of point 2 can be chosen to satisfy the linearised
constraint equations up to error terms which are $o(r^{\ell})$ in
a $C^k(B(p_0,r))$ norm, for small $r$.
\end{enumerate}
\end{Theorem}
%
\proof The proof of points 1 and 2 follows closely that of points
1 and 2 of Theorem~\ref{TnoCKV}. Given any constant $\alpha$, the
set $\mcJ_5$ of \eq{defRtwo} is replaced by the
\bean
&&\{\mbox{the set of fourth jets of $g_{ij}$ and of third jets of
$K_{ij}$ at $p$ that one obtains }
\\&& \phantom{\{} \mbox{as $g_{ij}$ varies in the set of all
Riemannian metrics in normal coordinates}\nonumber
\\&& \phantom{\{} \mbox{near $p$  and as $K_{ij}$
varies  in the set of all symmetric tensors}\nonumber \\&&
\phantom{\{} \mbox{with trace equal to a prescribed constant
$\alpha $}\}\;.\eeal{defRtwokid} The intermediate elements of the
proof are provided by Lemma~\ref{L3.1} and the first part of
Proposition~\ref{PnoKIS}. The variations of $g_{ij}$ and of the
trace-free part of $K_{ij}$ can be chosen to be polynomials
multiplied by a smooth cut-off function, and are therefore smooth.
One can then adjust the trace part of $K_{ij}$ to achieve
$K_\epsilon=\alpha$. We further note that the non-vanishing of
some derivative of $Q$ follows immediately from the fact that
$Q(p)$ is a polynomial, when viewed as a function depending upon
the jets of $g_{ij}$ and $K_{ij}$ in normal coordinates at $p$.
Further details are left to the reader.

In order to prove point 3, for $r>0$ it is useful to introduce the
following set:
\beaa \mcW_{\ell+k}&=&\{ \mbox{ jets at $p$ of order
$(\ell+k+1,\ell+k+2)$ of } \ (K_{ij},g_{ij}) \ \mbox{ such that }
\\ &&\ \rho=o(|\xi|^{\ell+k})\;, \ J=o(|\xi|^{\ell+k}) \
\mbox{ in } \ B(0,r)\}\;.\eeaa Here $\xi$ are supposed to be
geodesic coordinates near $p$ in the metric $g_{ij}$.
Equivalently, if $(K_{ij},g_{ij})\in C^{k+\ell+1}\times
C^{k+\ell+2}$ has jets in $\mcW_{\ell+k}$, then we have
\bel{loccon} D^{\alpha}\rho (p)=0\;,\ D^{\alpha}J(p)=0\;,\quad 0\le
|\alpha|\le \ell+k\;,\ee where the $\alpha=(i_1\ldots i_j)$'s are
multi-indices, with $|(i_1\ldots i_j)|=i_1+\ldots+i_j$. Elements
of $\mcW_{\ell+k}$ can be \emph{uniquely} parameterised as
follows: Taylor expanding $g_{ij}$ and $P_{ij}:=K_{ij}- K g_{ij}$
in geodesic coordinates around $p$, one can write
\bel{ggeo}g_{ij}=\delta_{ij}+\sum_{2\le|\alpha|\le
\ell+k+2}h_{ij\alpha}\xi^\alpha+O(|\xi|^{\ell+k+3})\;,\quad
\mbox{with}\ h_{i(j_1\ldots j_p)}=0\;,\ee
\be P_{ij}=\zP_{ij}+\sum_{1\le|\alpha|\le
\ell+k+1}P_{ij\alpha}\xi^\alpha+O(|\xi|^{\ell+k+2})\ee (see,
\emph{e.g.},\/~\cite{Thomas} for a justification of the last
condition in \eq{ggeo}). Then \eq{loccon} can be solved by
induction as follows: \eq{loccon} with $|\alpha|=0$ gives
\beaa&\sum_{i,j} \left( h_{ijij}- h_{jjii}\right)
=\sum_{i,j}\zP^2_{ij}- (\sum_i\zP_{ii})^2+2\Lambda\;,&\\& \sum_i
P_{ijj}= 0\;.&\eeaa For any given $\zP_{ij}\in \R^{m_0}:=\R^6$ the
first equation defines an affine subspace isomorphic to $\R^{n_2}$
for some $n_2$, in the vector space of second Taylor coefficients
$h_{ijkl}$. The second equation defines a linear subspace
isomorphic to $\R^{m_1}$ in the space of $P_{ijk}$'s, for some
$m_1$. To understand \eq{loccon} with $|\alpha|\ge 1$ we will need
the following:

\begin{Proposition}
\label{PapP} Let $k\in\N$, and suppose that $\dim M=n\ge 2$.
\begin{enumerate} \item For every
$J_i=J_{ij_1\ldots j_k}\xi^{j_1}\ldots \xi^{j_k}$ and $p=
p_{j_1\ldots j_{k+1}}\xi^{j_1}\ldots \xi^{j_{k+1}}$ there exists
$P_{ij}=P_{ijj_1\ldots j_{k+1}}\xi^{j_1}\ldots \xi^{j_{k+1}}$,
symmetric in $i$ and $j$, such that
$$\sum_i \partial_j P_{ij}= J_i\;, \quad \sum_i P_{ii}=p\;.$$
\item
For every  $f=f_{j_1\ldots j_k}\xi^{j_1}\ldots \xi^{j_k}$ there
exists $h_{ij}=h_{ijj_1\ldots j_{k+2}}\xi^{j_1}\ldots
\xi^{j_{k+2}}$, symmetric in $i$ and $j$, with $h_{i(jj_1\ldots
j_{k+2})}=0$, such that
$$\sum_{i,j} \left(\partial_j\partial_i h_{ij}-\partial_i\partial_i h_{jj}\right)=f\;.$$
\end{enumerate}
\end{Proposition}

\proof
 Consider a system of linear PDEs \bel{pde} Pu=I\;,\ee with constant
 coefficients, of  order $p$, which can be written in the Cauchy-Kowalevska form
with respect to a coordinate $z$.  We claim that if  $I$ is a
polynomial
 of order $l$, then there exists a solution of \eq{pde} which is a
 polynomial of order $l+p$. In order to see that, we note that \eq{pde} determines, at
 $z=0$, the $z$-derivatives of $u$ of order greater than or equal to $p$ as polynomials in the remaining
 variables. So choosing zero Cauchy data on $\{z=0\}$ one obtains
 a polynomial solution in $z$ with polynomial coefficients, hence a polynomial. If
 $P$ is homogeneous of order $p$, and if  $ I $ is
 in addition homogenous of order $l$, then the above solution is a
 homogeneous
 polynomial of order $l+p$.

 In order to prove point 1, we make the ansatz $$P_{ij}=\partial_i W_j + \partial_j W_i
 +
\frac{1}{n}\left(p-2\sum_\ell \partial _\ell W_\ell\right)
\delta_{ij} \;,$$ which leads to a homogeneous second order
elliptic system for $W$, and the above argument applies.

In order to prove point 2, we  first make the ansatz $
h_{ij}=\frac{1}{n}h_{ll}\delta_{ij}$, solve the resulting Poisson
equation in the class of  homogeneous polynomials as described
above, and introduce a metric $g_{ij}=\delta_{ij}+h_{ij}$. In
geodesic coordinates $y^i$ the metric $g_{ij}$ will have an
expansion with some new coefficients satisfying the symmetry
condition in~\eq{ggeo}~\cite{Thomas}. One has
$y^i=\xi^i+O(|\xi|^{k+3})$, which implies that the polynomial
obtained from the $y$--Taylor coefficients of $g_{ij}$ of order
$k+2$ provides the desired $h_{ij}$. \qed

Proposition~\ref{PapP} shows that~\eq{loccon} can be used to
inductively determine higher order Taylor coefficients
$h_{ij\alpha}$ and $P_{ij\beta}$ in terms of lower order ones, as
well as in terms of some free $P$--coefficients in
$\R^{m_{|\beta|}}$, for some ${m_{|\beta|}}\in \N$, and some free
$h$-coefficients in $\R^{n_{|\alpha|}}$, for some
${n_{|\alpha|}}\in \N$. It follows in particular that
$\mcW_{\ell+k}$ is diffeomorphic to $\R^{N_{\ell+k}}$, for some
$N_{\ell+k}\in \N$.

For solutions $(K_{ij},g_{ij})$ of the constraint equations, the
polynomial $Q[K_{ij},g_{ij}](p)$ can be expressed as a polynomial
of $(K_{ij},g_{ij})$-jets at $p$ of order $(2,3)$, call this
polynomial $\tilde Q$. Since the $\mcW_{\ell+k}$'s are included in
each other in the obvious way, $\tilde Q$  can actually be viewed
as a function defined on $\mcW_{\ell+k}$ which depends only on
those coefficients which parameterise $\mcW_{1}$. The pair
$(K_{ij},g_{ij})$ constructed in Proposition~\ref{PnoKIS} has jets
in $\mcW_1$, which shows that $\tilde Q$ is non-trivial on
$\mcW_1$. It then follows, as in the proof of
Theorem~\ref{TnoCKV}, that any jets in $\mcW_1$ can be
$\epsilon$-perturbed so that $\tilde Q(p)$ does not vanish on the
perturbed jet, with the jets of the perturbation belonging to
$\mcW_{\ell+k}$; by analyticity some of the derivatives of $\tilde
Q$ with respect to its arguments will not
vanish at $p$.

It should be clear from \eq{loccon} that the perturbed solution
satisfies the properties described in the statement of point 3 of
Theorem~\ref{TnoKIDs}. \qed

\section{Riemannian metrics without static KIDs near a point}
\label{Snoskids}

 An interesting class of
initial data is provided by the \emph{time-symmetric} ones,
$K\equiv 0$. In this case the KID equations \eq{k1}-\eq{k2}
decouple, with $X$ in \eq{k1} being simply a Killing vector field
of $g$. It remains to analyse the equation for $N$,
\begin{equation}
 \label{third} D_iD_jN = NR_{ij} + \Delta N g_{ij}\;.
\end{equation}
A solution of \eq{third} will be called  a \emph{static} KID, and
the set of static KIDs on a set $\myOmega$ will be denoted by
$\mcN(\myOmega)$.  (The origin of the adjective ``static" will be
clarified shortly.) Since time-symmetric initial data are
non-generic amongst all initial data, the results of the previous
section do not say anything about non-existence of static KIDs,
and separate treatment is required.

Taking  the trace of (\ref{third}) one obtains, in dimension  $n$
\begin{equation}
\label{fourth} \Delta N = -\frac{1}{n-1}NR\;,
\end{equation}
so that \eq{third} can be rewritten as
\begin{equation}
 \label{thirda} D_iD_jN = N(R_{ij} - \frac{1}{n-1}g_{ij}R)\;.
\end{equation}
Calculating $D^j$ of (\ref{thirda}) and commuting derivatives one
is led to (recall that the Einstein tensor is divergence-free)
\begin{equation}
\label{fifth} ND_iR =0\;.
\end{equation}
Since the zero-set of solutions of \eq{third} has no interior
except if $N\equiv 0$, we conclude that existence of non-trivial
static KIDs implies that $R$ is constant. It follows that a
non-trivial solution of \eq{thirda} does indeed correspond to
initial data for a static solution of the vacuum Einstein
equations with a cosmological constant. Further, one immediately
obtains that generic $C^2$ metrics have no static KIDs: it
suffices to vary the metric so that the scalar curvature is not
constant.

{}From now on we assume $\dim M = 3$. In order to prepare the
proof, that generic metrics with fixed constant value of scalar
curvature have no static KIDs, we consider a metric $g$ with Ricci
tensor at $p$ equal to
\begin{equation}
\label{first} \mathring R_{ij} = Ax_ix_j + By_iy_j + Cz_iz_j\;,
\end{equation}
where we assume that
$(A-B)(A-C)(B-C) \neq 0$, %
and we further suppose that
\begin{equation}
\label{second} D_{[i}\mathring R_{j]l} =  \alpha x_{[i}y_{j]}z_l +
\beta z_{[i}x_{j]}y_l + \gamma y_{[i}z_{j]}x_l - \frac{1}{6}
(\alpha + \beta + \gamma)\epsilon_{ijl}\;,
\end{equation}
with $(\alpha, \beta,\gamma) \ne 0$. We also impose the condition
that
\begin{equation}
\label{sixth} D_i \mathring R =0\;.
\end{equation}
Taking a curl of (\ref{third}) we infer that
\begin{equation}
\label{seventh} (2R_{j[l} - Rg_{j[l})D_{i]}N + g_{j[l} R_{i]k} D^k
N = N D_{[l}R_{i]j}\;.
\end{equation}
The left-hand side of (\ref{seventh}), with
\begin{equation}
\label{eighth} D_i \mathring N = a x_i + b y_i + c z_i\;,
\end{equation}
takes the form
\begin{eqnarray}
\label{ninth}
\lefteqn{x_jx_{[l}[y_{i]}b(A-C)+ z_{i]}c(A-B)]}&& \nonumber \\
&&+y_jy_{[l}[x_{i]}a(B-C) + z_{i]}c(B-A)] \nonumber \\&&
+z_jz_{[l}[x_{i]}a(C-B) + y_{i]}b(C-A)]\;.
\end{eqnarray}
Since no terms with this index structure occur in (\ref{second})
we obtain that $D_i \mathring N$ vanishes, and using
(\ref{seventh}) allows us to finally conclude that
\begin{equation}
\label{tenth} \mathring N = D_i \mathring N = 0\;.
\end{equation}
 The arguments of proof of Proposition~\ref{PnoCKV2}
apply and provide existence of a metric $g_{ij} = \delta_{ij} +
h_{ij}$ satisfying (\ref{first}) and (\ref{second}). Clearly
$A+B+C$ can be chosen so that $\mathring R$ has any prescribed
value. Now, we can multiply $g_{ij}$ by $1+\alpha$, where $\alpha$
is a homogeneous third order polynomial chosen  so that $D_i
\mathring R$ is zero. By conformal invariance this does not change
the value of $\mathring B_{ijk}$, hence of \eq{second} (compare
\eq{L}-\eq{cotton}), and does not change the value of $\mathring
R$ either. A repetition of the remaining arguments of
Section~\ref{SnoKIDSnearp}, with $K_{ij}$ there set to zero,
gives:

\begin{Theorem}
\label{TnoKIDsstatic} Let $(M,g)$ be a Riemannian manifold with
$g\in C^{k}$, $k\ge 2$. A necessary condition for a non-trivial
$\mcN(\myOmega)$ is that the scalar curvature of $g$ be constant
on $\myOmega$. Further, in dimension three, and for $k\ge 3$, the
following hold:
\begin{enumerate}
\item There exists a non-trivial homogeneous invariant polynomial $\Qpoly
 [g]:=\Qpoly (\Ric,D\Ric)$
 such that if $$\Qpoly (\Ric,D\Ric)(p)\ne 0$$ at a
point $p\in M$,  for a metric for which the gradient of the scalar
curvature vanishes at $p$, then there exists a neighborhood
$\mcO_p$ of $p$ for which there exist no non-trivial static KIDS
on any open subset of $\mcO_p$.
\item Let $\myOmega$ be a  domain in $M$, and let $p\in \myOmega$.
There exists a variation $ \delta g\in {C^{\infty}(\myOmega)}$,
compactly supported in $\myOmega$, such that $\Qpoly [
g+\epsilon\delta g](p)\ne 0$ for all $\epsilon $ small enough. If
$g\in C^{k+\ell+2}$, $\ell\ge 0$, has constant scalar curvature,
then the variation above can be chosen to have the same scalar
curvature up to error terms which are $o(r^{\ell})$ in a
$C^k(B(p_0,r))$ norm, for small $r$.
\end{enumerate}
\end{Theorem}

\section{Results in general dimensions, with  non-explicit orders of differentiability}\label{Spwc}

The results obtained so far did require rather unpleasant,
tedious, and lengthy calculations, and we will present here an
argument which avoids those. The draw-back is that one does not
obtain an explicit statement on the number of derivatives
involved. However, non-genericity of KIDs is obtained in higher
dimensions. Further, the proof below generalises immediately
\emph{e.g.}\/ to the Einstein-Maxwell equivalent of the KID
equations, the details are left to the reader.

The starting point of the analysis in this section is the
following result (recall that $n=\dim M$):

\begin{Lemma} \label{Lanex}
\begin{enumerate}
\item \label{Lanex1}For any $n\ge 2$ and for any signature there exists a real analytic compact
pseudo-Riemannian manifold $(M,g)$ without local Killing vectors.

\item\label{Lanex2n} For any $n\ge 3$ there exists a real analytic
compact simply connected Riemannian manifold $(M,g)$ without local
conformal Killing vectors.
\item\label{Lanex3} For any $n\ge 3$, $\Lambda \in \R$, $\tau \in \R$
there exists a real analytic vacuum initial data set $(M,g,K)$,
with cosmological constant $\Lambda$, with $\tr_gK=\tau$, and
without local KIDs.
\item \label{Lanex4} For any $n\ge 3$ and $\Lambda \in \R$,
there exists a real analytic Riemannian or Lorentzian manifold
$(\mcM,g)$, with $\dim \mcM=n+1$, satisfying the vacuum Einstein
equations with cosmological constant $\Lambda$ and without local
Killing vectors.\end{enumerate}
\end{Lemma}

\begin{Remark}
The main point of the Lemma is to construct \emph{one single}
example in each category listed. However, our argument makes it
clear that there are actually lots of examples. For instance, the
proof below shows that in point 1 for any analytic manifold $M$
which is simply connected, compact,  with $\dim M\ge 2$ one can
find a $g$ with the required properties.
\end{Remark}

\begin{Remark}\label{RLanex3}
If
\bel{taulamc2}
\tau^2 \ge  \frac
 {2n}{(n-1)}
 \Lambda\;,\ee
 then one can find an $M$ as in point 3 which is  compact (without
 boundary).  The proof in the strict inequality case is given below. If the inequality in \eq{taulamc} is an
equality, in the proof below one should instead choose
$(M,\gamma_0)$ to be any real analytic compact Riemannian manifold
of positive Yamabe class.  The monotone iteration scheme for
solving the Lichnerowicz equation can then be handled by an
argument in~\cite{Jimconstraints}. In that last reference only
dimension three is considered, but the proof applies in any
dimension.\end{Remark}

\proof \ref{Lanex1}: Let $M$ be any simply connected, compact,
analytic Riemannian manifold with $\dim M\ge 2$, let $p\in M$ and
let $g_0$ be any smooth Riemannian metric on $M$ such that the
polynomial $P_n=P_n[g]$ of theorem \ref{TnoKV} does not vanish at
$p$. We need analytic approximations of $g$, for example for $0\le
t< \epsilon$ we can let $g_t$ be the family of metrics obtained by
evolving $g_0$ using the Ricci flow, then the metrics $g_t$ are
indeed real analytic for $t>0$. By continuity, reducing $\epsilon$
if necessary,  we will have $P_n[g_t](p)\ne 0$, hence $g_t$ will
have no Killing vectors in a neighborhood of $p$. Now,  a
theorem\footnote{We note that in~\cite{Nomizu} a Riemannian metric
is assumed, but the proofs given there apply to any signature.} of
Nomizu~\cite{Nomizu} shows that on a simply connected analytic
manifold every locally defined Killing vector extends to a
globally defined one. This implies that for $0<t <\epsilon$ the
metrics $g_t$ have no Killing vectors on any open subset of $M$.

\bigskip

\ref{Lanex2n}: For $n=3$ this follows from Theorem~\ref{TnoCKV}.
For any $n\ge 3$ one can argue as follows: Let $M$ be any compact
real analytic manifold of dimension not less than three.
By~\cite{Lohkamp:Ricci} there
exists on $M$ a metric $g$ with strictly negative Ricci curvature.
It is well known that such metrics do not have non-trivial
conformal Killing vectors, we recall the proof for completeness:
from \eq{conf} with $2/3$ replaced by $2/n$ it follows that
\beq \label{cor1n} D_i
D_{j}X_{k} = - R_{jki}{}^{l} X_l + \frac{1}{n}\left( \pdivX
_{i}g_{jk}+\pdivX _{j}g_{ik}-\pdivX _{k}g_{ij}\right) \;, \eeq
hence
\beq\label{cor1n2} \Delta X_{k} = - R_{k}{}^{l} X_l -(1 - \frac{2}{n})\pdivX _{k}
\;. \eeq Multiplying by $X^k$ and integrating over $M$ one finds
(recall that $\varphi= \mathrm{div} X$)
$$\int_M |DX|^2 - \Ric(X,X) + (1 - \frac{2}{n})\varphi^2 = 0\;,$$
so that $X\equiv 0$ if $\Ric <0$. Approximating $g$ by
real-analytic metrics $g_t$, $g_t\to g$ as $t\to 0$, one will have
no conformal Killing vectors for $g_t$ when $t$ is small enough by
Proposition~\ref{Pclosed}. It then follows from
Theorem~\ref{Textend}, Appendix~\ref{Sltg}, that the $g_t$'s will
have no local conformal Killing vector field either.

\bigskip

 \ref{Lanex3} and \ref{Lanex4}: {We start by noting that in
 dimension $n=3$,  an example of initial data as in point~\ref{Lanex3} can be obtained
using vacuum Robinson-Trautman space-times with cosmological
constant $\Lambda$ (\emph{cf., e.g.,}~\cite{Bicak:podolsky}).
Because of the parabolic character of the Robinson-Trautman
equation, those metrics are always analytic away from the initial
data surface. Further, if the initial metric $h_0$ on $S^2$ used
in the Robinson-Trautman equation has no continuous global
symmetries, then it follows from Proposition~\ref{Pclosed} that
the evolved metrics $h_t$ will not have any continuous global
symmetries either, at least for $t$ small enough. It is clear that
the resulting four-dimensional metric $\fourg$ will then have no
globally defined Killing vectors except the zero one. The
non-existence of local Killing vectors follows then from Nomizu's
theorem~\cite{Nomizu}. Finally, the initial data set of point
\ref{Lanex3} can be obtained as that induced by $\fourg$ on any
hypersurface with $\tr_g K=\tau$ in $\mcM$; such hypersurfaces can
be obtained by solving a Dirichlet problem for the CMC equation on
the boundary of a sufficiently small spacelike
three-ball~\cite{bartnik:variational}.

 In any case,
whatever $n\ge 3$ one can proceed as follows: consider, first,
$\tau$ such that the inequality in \eq{taulamc2} is strict, let
$(M,\gamma_0)$ be any real analytic compact Riemannian manifold of
negative Yamabe class. Let $L_0$ be any non-zero,
$\gamma_0$-transverse and traceless tensor on $M$; such tensors
exist by~\cite{BEM}. For $t\in [0,\epsilon)$ let $L_t$ be a family
of analytic symmetric $\gamma_0$-trace free
tensors converging to $L_0$. For example, 
$L_t$ can be obtained from $L_0$ by heat flow using any analytic
metric on $M$, and removing the $\gamma_0$ trace. Using the
conformal\footnote{Since the inequality in \eq{taulamc} is strict,
a small and a large constant provide barriers for the monotone
iteration scheme.} method~\cite{Jimconstraints} with seed fields
$(\gamma_0,L_t)$ one obtains a family of real analytic vacuum CMC
initial data sets $(g_t,K_t)$ with cosmological constant
$\Lambda$. Since $\gamma_0$ has no global conformal Killing
vectors, $g_t$ will have no global Killing vectors. Now,
$\tr_{g_t}K_t=\tau $ is a constant, which implies (see
Remark~\ref{RP3} below) that any global KIDs for $(g_t,K_t)$ are
of the form $(N=0,Y)$, where $Y$ is a Killing vector of $g_t$,
therefore none of the $(g_t,K_t)$'s has global KIDs. In the
Lorentzian case we let $(\mcM,\nponeg_t)$ be the maximal globally
hyperbolic vacuum development of $(M,g_t,K_t)$, then $\mcM$ is
diffeomorphic to $\R\times M$ (hence simply connected), and
$\nponeg_t$ is analytic by~\cite{AlinhacMetivier}. In the
Riemannian case we let $\mcM$ be any simply connected and
connected neighborhood $\mcU_t$ of $M\times \{0\}$ in $M\times
(-1,1)$, chosen so that there exists a vacuum metric $\nponeg_t$
on $\mcU_t$ with Cauchy data $(g_t,K_t)$ on $M\times \{0\}$,
obtained from the Cauchy-Kowalewska theorem. Suppose that there
exists an open nonempty subset $\myOmega_t\subset M$ such that
$\mcK_t(\myOmega_t)\ne \{0\}$, where $\mcK_t$ denotes the set of
KIDs with respect to $(g_t,K_t)$, then a standard
argument~\cite{CollJMP}\footnote{Compare the proof of
\cite[Theorem~2.1.1]{SCC}; the Cauchy-Kowalewska theorem should be
invoked for solvability of Eq.~(2.1.5) there, or for uniqueness of
solutions of Eq.~({2.1.7}) there.}, using the Cauchy-Kowalewska
theorem, shows that there exists a non-trivial Killing vector $X$
in a neighborhood of $\myOmega_t$ in $\mcM$. By Nomizu's
theorem~\cite{Nomizu} $X$ extends to a globally defined Killing
vector on $\mcM$, hence $(M,g_t,K_t)$ has a globally defined KID,
a contradiction. Thus there are no local KIDs on $(M,g_t,K_t)$,
and $(\mcM,\nponeg_t)$ is a vacuum metric without local Killing
vectors. This proves point 4, as well as Remark~\ref{RLanex3} in
the case of a strict inequality there. To prove point 3 for the
remaining values of $\tau$ one  spans~\cite{bartnik:variational},
within the Lorentzian solution $\mcM$ just constructed, a CMC
hypersurface of prescribed $\tau=\tr_g K$ on the boundary of a
small spacelike ball. The data induced on the resulting CMC
hypersurface provide the desired initial data set. \qed

We continue with the question of generic non-existence of KIDs, it
should be clear that an identical argument applies to conformal
Killing vectors (compare \cite[Equation~(1.15)]{BCEG}), or to
Killing vectors. Let $(g,K)$ be any vacuum analytic initial data
on a simply connected manifold $M$ which have no global KIDs. As
explained above, it follows from a theorem of
Nomizu~\cite{Nomizu}, that such an initial data set will not have
any local KIDs. Let $r(p)\in \R^M$ be as in Lemma~\ref{L3.1}, for
some appropriate $M$, and for $\alpha\in \N^n$, 
let us write
\bel{ksys} D_{\alpha}r=P_\alpha r\ee for the linear system of equations obtained by
calculating the $(k+1)$-st derivatives of $r$ by differentiating
\eq{k1}-\eq{k2} $|\alpha|$ times, and replacing the lower order
derivatives that arise in the process by their values already
calculated from the previous equations.   Let us write $L_k r = 0$
for the system of equations that arise from first-order
integrability conditions of the system \eq{ksys} with $|\alpha|\le
k$. Choose some orthonormal frame, then $L_k$ can be identified
with a $N_k\times M$ matrix, for some $N_k$, with entries built
out of the extrinsic curvature tensor $K$, of the Riemann tensor,
and of their derivatives. Let $Q_k$ denote the sum of squares of
determinants of all $M\times M$ sub-matrices of $L_k$. Then the
equation $L_kr=0$ admits a non-trivial solution if and only if
$Q_k=0$. Suppose that there exists $r_0$ such that $L_kr_0=0$ for
all $k$. One can then use \eq{ksys} to calculate all the jets of
$r$ with initial value $r_0$ at $p$ so that the Killing equations
are satisfied to infinite order at $p$ by a formal solution
determined by those jets. To show convergence of the resulting
Taylor series one can proceed as follows: let $x^i\in
[-\epsilon,\epsilon]^n$ be local analytic coordinates around
$p=\vec 0 $, we can solve the linear equation
$$\frac{\partial r}{\partial x^1} = P_1r$$ along the path $[-\epsilon,\epsilon]\ni
x^1\to(x^1,0,\ldots,0)$, with initial data $r_0$ at the origin,
obtaining an analytic solution there. We can use the function so
obtained as initial data for the equation
$$\frac{\partial r}{\partial x^2}= P_2r$$ to obtain an analytic solution on
$[-\epsilon,\epsilon]^2\times\{0\}\times\ldots\times \{0\}$. An
inductive repetition of this procedure provides an analytic
solution on $[-\epsilon,\epsilon]^n$ of the
equation
$$\frac{\partial r}{\partial x^n} = P_nr\;,$$ such
that  $$\mbox{the equation  } \ \frac{\partial r}{\partial x^k}=
P_kr\ \mbox{holds on
$[-\epsilon,\epsilon]^k\times\underbrace{\{0\}\times\ldots\times
\{0\}}_{\mbox{\scriptsize  $n-k$ factors}}$.}$$ By choice of $r_0$
the analytic functions $L_kr$ have all derivatives vanishing at
the origin, hence they vanish on $[-\epsilon,\epsilon]^n$.
Standard arguments imply that the function $r$ so obtained
provides an analytic solution of the KID equations in a
neighborhood of $p$. This gives a contradiction with the fact that
$(g,K)$ has no local KIDs near $p$. Therefore there exists $k$
such that $Q_k$ is non-zero for the initial data set under
consideration. This $Q_k$ provides the non-trivial polynomial
needed in Theorem~\ref{TnoKIDs}. When the metric involved is
Riemannian we can integrate $Q_k$, viewed as a function on the
frame bundle, over the rotation group to obtain an invariant
polynomial. We have therefore proved:

\begin{Theorem}
\label{Tgen} Theorem~\ref{TnoKIDsstatic} remains valid in any
dimension, with an invariant polynomial that depends upon some
dimension-dependent number  $k$ of derivatives of $g$. Similarly
Theorem~\ref{TnoKIDs} remains valid in any dimension, for some
polynomial that depends upon $k+1$ derivatives of $g$ and $k$
derivatives of $K$, for some dimension-dependent number $k$. In
dimension $n\ge 4$ Theorem~\ref{TnoKV} remains valid with a
polynomial which depends upon some dimension-dependent number $k$
of derivatives of the Weyl tensor. Finally, Theorem~\ref{TnoCKV}
remains valid in any dimension $n\ge 3$ with a polynomial that
depends upon some dimension-dependent number of derivatives of the
Riemann tensor.
\end{Theorem}

\proof The only statement which, at this stage, might require
justification is the extension of Theorem~\ref{TnoKV}: this result
follows from point 4 of Lemma~\ref{Lanex}, as the polynomial
obtained in that case by the proof above depends only upon the
Weyl tensor.

\hfill\qed

\section{{}From  approximate linearised solutions to small vacuum perturbations}
\label{Sexact}

The perturbation results of the previous sections can be used to
prove non-genericity of KIDs when no restrictions on $\rho$ and
$J$ are imposed. They also apply if, \emph{e.g.},\/ a
\emph{strict} dominant energy condition $\rho > |J|$ is imposed,
for then a sufficiently small perturbation of the data will
preserve that inequality. However, some more work is needed when
vacuum initial data are considered, and this is the issue
addressed in this section.

Let $\myOmega \subset M$ be open and connected, and let
$\mcK(\myOmega)$ denote the set of KIDs defined on $\myOmega$;
each $\mcK(\myOmega)$ is a finite dimensional, possibly trivial,
vector space. If $\myOmega'\subset \myOmega$ we have the natural
map
$$i_{\myOmega'}: \mcK(\myOmega)\to \mcK(\myOmega')\;,$$
with $i_{\myOmega'}(x)$ being defined as the restriction to
$\myOmega'$ of the KID $x\in \mcK(\myOmega)$. A local KID
vanishing on an open subset vanishes throughout the relevant
connected component of its domain of definition, which shows that
$i_{\myOmega'}$ is injective.

We denote by $B(p,r)$ the open geodesic ball of radius $r$, and
for $a<b$ we set $\Gamma_p(a,b):= B(p,b)\setminus
\overline{B(p,a)}$.

 We will need the following result:

 \begin{Proposition}
 \label{Prest} For every $p\in M$ and  $\R\ni r> r_1>0$ there exists $0<r_2<r_1$ such that
$$i_{\Gamma_p(r_2,r)}: \mcK(B(p,r))\to \mcK(\Gamma_p(r_2,r))$$
is  bijective.
 \end{Proposition}

 The proof rests on the following lemma:
 \begin{Lemma}
 \label{Lrest} For every $p\in M$ and $r_1>0$ there exists  $\sigma \in (0,1)$ such that
$$i_\sigma: \mcK(B(p,r_1))\to \mcK(\Gamma_p(\sigma r_1,r_1))$$
is  bijective.  Here $i_\sigma$ denotes $i_{\Gamma_p(\sigma
r_1,r_1)}$.
 \end{Lemma}

\proof As already pointed out, injectivity always holds. Suppose
that surjectivity fails, then for
every $\sigma \in (0,1)$ there exists a KID
$x_\sigma\in\mcK(\Gamma_p(\sigma r_1,r_1))$ such that
$x_\sigma\not\in i_\sigma(\mcK(B(p,r_1)))$. Choose
any scalar product $h$  on $\mcK(\Gamma_p(r_1/2,r_1))$. For
$\sigma< 1/2$ without loss of generality we can assume that the
restriction $\hat x_\sigma$ of  $x_\sigma $ to
$\Gamma_p(r_1/2,r_1)$ is $h$-orthogonal to the image of $i_{1/2}$,
and that $h(\hat x_\sigma,\hat x_\sigma)=1$. Since
$\mcK(\Gamma_p(r_1/2,r_1))$ is finite dimensional there exists a
sequence $\sigma_i\to 0$ such that $\hat x_{\sigma _i}$ converges
to some $\hat x_0$, with $h(\hat x_0,\hat x_0)=1$. Further $\hat
x_0$ is $h$-orthogonal to $i_{1/2}(\mcK(B(p,r_1)))$. It should be
clear from \eq{k2}-\eq{k1b} that for $i$ such that $\sigma >
\sigma_i$, the sequence of KIDs on $\Gamma_p(\sigma r_1, r_1)$
obtained by restricting $x_{\sigma_i} $ to $\Gamma_p(\sigma r_1,
r_1)$ converges, and defines a non-trivial KID which restricts to
$\hat x_0$ on $\Gamma_p(r_1/2,r_1)$, with the limit being
independent of $\sigma$ in the obvious sense. This shows that
there exists a KID $x_0$ defined on $B(p,r_1)\setminus\{p_1\}$
such that $\hat x_0$ is the restriction of $x_0$ to
$\Gamma_p(r_1/2,r_1)$. But \eq{k2}-\eq{k1b} further shows that
$x_0$ can be extended to a KID defined on $B(p,r_1)$, still
denoted by $x_0$. It follows that $\hat x_0 = i_{1/2}(x_0)$, which
contradicts orthogonality of $\hat x_0$ with the image of
$i_{1/2}$. \qed

\noindent {\sc Proof of Proposition~\ref{Prest}}: Let $r_2=\sigma
r_1$, with $\sigma$  given by Lemma~\ref{Lrest}. Every KID on
$\Gamma_p(r_2,r)$ induces, by restriction, a KID on
$\Gamma_p(r_2,r_1)$, therefore $\dim \mcK(\Gamma_p(r_2,r)) \le
\dim \mcK(\Gamma_p(r_2,r_1))$. By Lemma~\ref{Lrest} we have $\dim
\mcK(\Gamma_p(r_2,r_1)) = \dim \mcK(B(p,r))$. Again by restriction
we have $\dim \mcK(B(p,r)) \le \dim \mcK(\Gamma_p(r_2,r))$, whence
the result. \qed

\begin{Corollary}
\label{Crest} Suppose that $\mcK(B(p,r))=\{0\}$. Then for any
$\epsilon>0$ there exists $\epsilon> r_1>0$ such that
$\mcK(\Gamma_p(r_1,r))=\{0\}$. \hfill $\Box$
\end{Corollary}

Recall that the {\em constraints map} has been defined by the
formula:
\be \label{1} \left(
\begin{array}{c}
J\\
  \\
\rho
\end{array}
\right) (K,g):= \left(
\begin{array}{l}
2(-\nabla^jK_{ij}+\nabla_i\;\tr  K)\\
  \\
R(g)-|K|^2 + (\tr  K)^2 -2 \Lambda
\end{array}
\right) \;. \ee The following is one of the key steps of the
proof:

\begin{Theorem} \label{Tal}
For $\ell\in\N$,  $\ell \ge 2$, $\alpha\in (0,1)$, $p\in M$,
$r,\eta>0$, let the symbol $P$ denote the linearisation of the
constraints operator \eq{1} at $(K,g)\in \left(C^{\ell+2,\alpha
}\times C^{\ell+2,\alpha }\right)(B(p,r))$, and let
 $x_\eta=(\delta K_\eta,\delta g_\eta)\in \left(C^{\ell+2,\alpha }\times
C^{\ell+2,\alpha }\right)(B(p,r))$ be an ``approximate solution"
of the linearised constraint equations defined on $B(p,r)$, in the
sense that:
 $$\|Px_\eta\|_{\left(C^{\ell+1,\alpha }\times C^{\ell,\alpha }\right)(B(p,r))}\le \eta\;.$$
 \begin{enumerate}
\item There exists a constant $C$ such that if $i_{\Gamma_p(\sigma r,r)}$
is surjective for some $\sigma\in (0,1/2]$, then there
 exists a solution $x\in \left(C^{\ell+2,\alpha }\times
C^{\ell+2,\alpha }\right)(B(p,r))$ of the linearised constraint
 equations supported in $B(p,r)$ such that
 \bel{Cconst}\|x-x_\eta\|_{\left(C^{\ell+2,\alpha }\times
C^{\ell+2,\alpha }\right)(B(p,\sigma r))}\le C \eta\;.\ee $x$ is
smooth if $(K,g)$ and $x_\eta$ are.
\item
For $\ell \ge 4$, for any $(K_0,g_0)$ in $\left(C^{\ell+2,\alpha
}\times C^{\ell+2,\alpha }\right)(B(p,r))$, and for any $r_0$ such
that $B(p,r_0)$ has smooth boundary, the constant $C$ can be
chosen independently of $\sigma\in (0,1/2]$, $(K,g)$, and $r$
satisfying $0<r\le r_0$, for all $(K,g)$ sufficiently close in
$\left(C^{\ell+2,\alpha }\times C^{\ell+2,\alpha
}\right)(B(p,r_0))$ to $(K_0,g_0)$.
 \end{enumerate}
 \end{Theorem}

 \begin{Remark}
 The restriction $\sigma\le 1/2$ is arbitrary, the argument
 applies with any $0<\sigma\le \sigma_0\in (0,1)$, with a constant
 in \eq{Cconst} depending perhaps upon $\sigma_0$.
 \end{Remark}

 \proof
We use the definitions and notation of~\cite{ChDelayHilbert}. In
particular if $\myOmega$ is a domain with smooth boundary, then
$$
 \Lambda^{s}_{k,\alpha}=\hbord^{s}_2\cap
\cbord^{s}_{k,\alpha}\;.$$ Roughly speaking, functions in that
space behave as $o(x^s)$ near the boundary $\{x=0\}$, with
derivatives of order $j$, $0\le j \le k$, being allowed to behave
as $o(x^{s-j})$. In particular if $s>k+\alpha$ then functions in
the space above are in $C^{k,\alpha}(\overline{\myOmega})$. We
will need the following
result~\cite[Proposition~6.5]{ChDelayHilbert}:

\begin{Proposition}
 \label{Psio} Suppose that $(K_0,g_0)\in \left(C^{k+2,\alpha}\times
 C^{k+2,\alpha}\right)(M)$, $k\ge 2$, $\alpha \in (0,1)$, and let $\myOmega\subset M$ be a domain with
 smooth boundary and compact closure.
For all $s\neq (n+1)/2,(n+3)/2$, the image of the linearisation
$P$, at $(K_0,g_0)$, of the constraints map, when defined on $
\left(\Lambda^{-s+1}_{k+2,\alpha}\times
\Lambda^{-s+2}_{k+2,\alpha}\right)(\myOmega)$, is%
$$
\begin{array}{c}\left\{(J,\rho)\in \Lambda^{-s}_{k+1,\alpha}\times
\Lambda^{-s}_{k,\alpha}\ \mbox{ such that } \
\langle(J,\rho),(Y,N)\rangle_{(L^2\oplus L^2)(\myOmega,d\mu_{g_0})}=0\;\right.\\
\left. \mbox{ for all } \  (Y,N)\in H^{s-n}_1\times H^{s-n}_2 \
\mbox{ satisfying } \  P^*(Y,N)=0\right\}\;. \end{array}$$ Further
$P^{-1}(0)\subset \Lambda^{-s+1}_{k+2,\alpha}\times
\Lambda^{-s+2}_{k+2,\alpha}$ splits. \qed
 \end{Proposition}

 The proof of point 1 of Theorem~\ref{Tal} will
proceed in two steps:

\medskip

{\noindent \sc Step 1;} We set $M:=B(p,r)$, $k=\ell$,
$(g_0,K_0)=(g,K)$, and we use Proposition~\ref{Psio} with $s=s_1$
for some $s_1<-1$. Now, for such $s$ the space $\maclK_0\subset
\maclK (B(p,r))$ above is the space of KIDs on $B(p,r)$ which
vanish at $S(p,r):=\partial B(p,r)$ together with their first
derivatives; but Equations~\eq{k2}-\eq{k1b} imply that there are
no such non-trivial KIDs. It follows that {$P$ is surjective},
with the splitting property being equivalent to the fact that
there exists a closed subspace $X\subset
\Lambda^{-s+1}_{k+2,\alpha}\times \Lambda^{-s+2}_{k+2,\alpha}$
such that the restriction of $P$ to $X$ is an isomorphism. This
shows that there exists
 $\hat x_\eta\in \Lambda^{{-s+1}}_{k+2,\alpha}\times
\Lambda^{{-s+2}}_{k+2,\alpha}$ satisfying
$$\|\hat x_\eta\|_{\Lambda^{{-s+1}}_{k+2,\alpha}\times
\Lambda^{{-s+2}}_{k+2,\alpha}}\le C \eta\;,$$ and $P(\hat
x_\eta)=-P( x_\eta)\ \Leftrightarrow \ P(x_\eta+\hat x_\eta)=0$.

\medskip

{\noindent \sc Step 2:}  Now, because $s=s_1<-1$, the correction
term $\hat x_\eta$ could be blowing-up near $S(p,r)$, while we
want a solution which vanishes there to rather high order. To
correct that, let $\varphi$ be any smooth non-negative function
which is identically one on $B(p,5r/8)$, and vanishes on
$\Gamma_p(3r/4,r)$, set
$$y_\eta= \varphi (x_\eta+\hat x_\eta)\;.$$ Then $P(y_\eta)$ is
supported in $\overline{\Gamma_p(5r/8,3r/4)}\subset \Gamma(\sigma
r, r)$. We now use Proposition~\ref{Psio} once again, with some
$s=s_2>\ell+3$, to find $\tilde x_\eta\in\left(C^{\ell+2,\alpha
}\times C^{\ell+2,\alpha }\right)(\Gamma_p(\sigma r,r))$, which
extends by zero both through $S(p,\sigma r)$ and through $S(p,r)$
in a $C^{\ell+2,\alpha }\times C^{\ell+2,\alpha }$ manner, such
that
$$P(\tilde x_\eta+ y_\eta)=0\ \Longleftrightarrow \ P(\tilde
x_\eta)= -Py_\eta=:z_\eta\;.$$ This will be possible if and only
if $z_\eta$ is orthogonal in $L^2(\Gamma_p(\sigma r,r))$ to
$\maclK_0(\Gamma_p(\sigma r,r))$, where now
$\maclK_0(\Gamma_p(\sigma r,r))$ coincides with the space of all
KIDs on $\Gamma_p(\sigma r,r)$. Let, thus, $w=(Y,N)\in
\maclK_0(\Gamma_p(\sigma r,r))$, by hypothesis there exists a KID
$\hat w$ defined on $B(p,r)$ such that $w$ is the restriction to
$\Gamma_p(\sigma r,r)$ of $\hat w$. We then have
$$\int_{\Gamma_p(\sigma r,r)} \langle w, P y_\eta\rangle =
\int_{B(p,r)} \langle \hat w, P y_\eta\rangle = \int_{B(p,r)}
\langle \hat P^*\hat w,  y_\eta\rangle =0\;.$$ Here the first and
the second equalities are justified because $Py_\eta$ is supported
in $\overline{\Gamma_p(\sigma r,3r/4)}$, while the last one
follows because, by definition of a KID, $P^*\hat w=0$. This
provides the desired $\tilde x_\eta$. Setting $x_\eta=\varphi( x+
\hat x_\eta) + \tilde x_\eta$, point 1 is proved.

To prove point 2, we first note that the value of $\sigma$ does
not affect the constant $C$, as that constant arises from step 1
of the proof of point 1: the perturbation $\tilde x_\eta$ from
step 2, which {\em could} depend upon $\sigma$, is supported away
from $B(p,\sigma r)$. The result is proved now by the usual
contradiction argument: Consider the map
\bel{map}
\pi_{\maclKzo }L_{\epsilon,x,x^{s-n/2}} :{\maclKzo }\cap
(\Lambda^{-s}_{k+3,\alpha}\times \Lambda^{-s}_{k+4,\alpha})
\longrightarrow {\maclKzo }\cap (\Lambda^{-s}_{k+1,\alpha}\times
\Lambda^{-s}_{k,\alpha}) \;,\ee  with $L_{\epsilon,x,x^{s-n/2}}$
being a regularised version, 
as in \cite{ChDelayHilbert},  of the map $L_{x,x^{s-n/2}}$ of
\cite[Section~5]{ChDelay}. \Eq{Cconst} will fail to hold only if
there exists a sequence of radii $r_n$ and data $(K_n,g_n)$ on
$B(p,r_n)$ near $(K_0,g_0)|_{B(p,r_n)}$, with KIDs $(Y_n,N_n)\in
{\maclKzo }$ such that
$$\|(Y_n,N_n)\|_{\Lambda^{-s}_{3,\alpha}\times
\Lambda^{-s}_{4,\alpha}}=1\ \mbox {and } \
\|L_{\epsilon,x,x^{s-n/2}}(Y_n.N_n)\||_{\Lambda^{-s}_{1,\alpha}\times
\Lambda^{-s}_{0,\alpha}}\le 1/n\;.$$ Consider an extracted
sequence, still denoted by $r_n$, converging to $r_\infty$. If
$r_\infty>0$, then $(K_0,g_0)|_{B(p,r_\infty)}$ would admit a KID
vanishing, together with its first derivatives, at
$S(p,r_\infty)$, a contradiction. On the other hand suppose that
$r_\infty=0$, introduce geodesic coordinates for the metrics
$(K_n,g_n)$ centred at $p$; this might lead to a loss of two
derivatives of the metric, so we increase the threshold on $\ell$
from two to four. Consider the sequence $(\tilde K_n,\tilde g_n)$
on $B(p,1)$ obtained by scaling up the ball $B(p,r_n)$ to
$B(p,1)$. Then $(\tilde K_n,\tilde g_n)$ converges to
$(0,\delta)$, where $\delta$ is the Euclidean metric on $B(p,1)$.
As before one obtains a contradiction because there are no  KIDs
vanishing, together with their first derivatives, on $S(p,1)$ for
$(K,g)=(0,\delta)$.

Smooth solutions can be obtained proceeding as above, but working
instead with exponentially-weighted rather than power-weighted
spaces. \qed

The main result of this section is the following (see
footnote~\ref{Fko}):

\begin{Theorem}\label{Tmain} Let $M$ be a compact
manifold with boundary, suppose that $\ell \ge \ell_0(n)$,
$\alpha\in (0,1)$ for some $\ell_0(n)$ ($\ell_0(3)=6$), and let
$(M,K,g)$ be a $C^{\ell,\alpha}\times C^{\ell,\alpha}$ vacuum
initial data set such that$$\mcK(M)=\{0\}\;.$$ For any $p \in
M\setminus
\partial M$ and for any $\epsilon>0$ there exists $r>0$ and an
$\epsilon$-small, in a $C^{\ell,\alpha}\times C^{\ell,\alpha}$
topology, vacuum perturbation $(K_\epsilon,g_\epsilon)$ of $(K,g)$
such that
$$
\mcK(\mcU)=\{0\} \ \mbox{ for all 
} \ \mcU \ \mbox{such that  } \ \mcU\cap  B(p,r)\ne \emptyset
\;.$$ Further, $(K_\epsilon,g_\epsilon)$ can be chosen to coincide
with $(K,g)$ in a neighborhood of $\partial M$.
\end{Theorem}

\proof  For definiteness in the proof we will assume $n=3$, for
$n>3$ in the argument below Theorem~\ref{TnoKIDs} should be
replaced by its higher-dimensional generalisation provided by
Theorem~\ref{Tgen}. If the polynomial $\Qpoly $ of point 1 of
Theorem~\ref{TnoKIDs} does not vanish at $p$, we let $r>0$ be
small enough so that $\Qpoly $ has no zeros on $B(p,r)$.
Otherwise, let $\delta x:=(\delta K, \delta g)$ be as in point 3
of Theorem~\ref{TnoKIDs} with $\ell=1$ and $k=3$. Let $\epsilon
>0$, as $\Qpoly$ is a polynomial we have
$$\Qpoly [x+\epsilon \delta x](p)= \epsilon^j (\Qpoly ^{(j)}[\delta x])(p) +O(\epsilon^{j+1})\;,$$
for some $j\ge 1 $ such that   $(\Qpoly ^{(j)}[\delta x])(p)\ne
0$. 
By Proposition~\ref{Prest}
for any $r>0$ there exists $\sigma_r\in (0,1)$ such that the
conditions of Theorem~\ref{Tal} are satisfied. We then have
$$\|P\delta x\|_{(C^{3,\alpha}\times C^{2,\alpha})(B(p,r))} \le
\|P\delta x\|_{(C^{4}\times C^{3})(B(p,r))} \le C_1 r\;,$$ and
Theorem~\ref{Tal} provides a solution $\delta \tilde x$ of the
linearised constraint equations supported in $B(p,r)$ such that
$$\|\delta x- \delta \tilde x\|_{C^{4,\alpha}(B(p,\sigma_rr))} \le C C_1 r\;. $$
Choosing $r$ small enough so that $CC_1r\le \epsilon$ one obtains
\beaa \Qpoly [x+\epsilon \delta \tilde x  ](p)= \epsilon^j (\Qpoly ^{(j)}[\delta
x +O(\epsilon)])(p) +O(\epsilon^{j+1}))= \epsilon^j (\Qpoly
^{(j)}[\delta x])(p) +O(\epsilon^{j+1}) \;.\eeaa
Since $\delta \tilde x $ satisfies the linearised constraint
equations and since $\mcK(M)=\{0\}$, it follows
from~{\cite[Theorem~5.6]{ChDelay}} together with the
regularisation technique from~\cite{ChDelayHilbert} that for
$\epsilon$ small enough we can find $\delta \hat x(\epsilon)$,
with $\|\delta \hat x(\epsilon)\|_{C^4(B(p,r))} \le C_2\epsilon
^2$, such that $x+\epsilon \delta \tilde x + \delta \hat
x(\epsilon)$ satisfies the vacuum constraint equations. Choosing
$\epsilon$ small enough so that $C_2\epsilon \le \epsilon^{1/2}$
we then obtain
\beaa
\Qpoly [x+\epsilon \delta \tilde x +\delta \hat x (\epsilon) ](p)
&=& \epsilon^j (\Qpoly ^{(j)}[\delta \tilde x +
O(\epsilon^{1/2})])(p) +O(\epsilon^{j+1})\\ &=& \epsilon^j (\Qpoly
^{(j)}[\delta x])(p) +O(\epsilon^{j+1/2}) \ne 0\eeaa
 for $\epsilon $ small enough. Replacing $r$ by a smaller number if necessary so that
 $\Qpoly [x+\epsilon \delta \tilde x +\delta \hat x (\epsilon) ]$ has no zeros on $B(p,r)$,
 the resulting data set has no
 KIDs in any subset of $B(p,r)$ by point 1 of Theorem~\ref{TnoKIDs}.

 The construction described so far leads to a perturbed initial data set
 which agrees with the starting one at $\partial M$ to arbitrarily
 high order. Consider, next, a collar neighborhood $N_s=\{p\in M: d(p,\partial M)< s\}$ of $\partial
 M$.
  Arguments as in Lemma~\ref{Lrest} show that $\mcK(M_s:=M\setminus
  N_s)=\{0\}$ for $s$ small enough. Applying the result already
  established to $M_s$ one obtains a perturbation which
  vanishes on $N_s$.
\qed

An identical proof, based on the results in
Section~\ref{Snoskids}, gives:
\begin{Theorem}\label{TmainRiem} Let $(M,g)$ be a $n$-dimensional $C^{\ell,\alpha}$
compact Riemannian manifold with boundary, suppose that $\ell \ge
\ell_0(n)$, $\alpha\in (0,1)$ for some $\ell_0(n)$, and suppose
that $g$ has constant scalar curvature $s$. Assume that there are
only trivial static KIDs, $$\mcN(M)=\{0\}\;.$$ For any $p \in
M\setminus
\partial M$ and for any $\epsilon>0$ there exists $r>0$ and an
$\epsilon$-small, in a $C^{\ell,\alpha}$ topology, perturbation
$g_\epsilon$ of $g$ with scalar curvature $s$ such that
$$
\mcN(\mcU)=\{0\} \ \mbox{ for all 
} \ \mcU \ \mbox{such that  } \ \mcU\cap  B(p,r)\ne \emptyset
\;.$$ Further, $g_\epsilon$ can be chosen to coincide with $g$ in
a neighborhood of $\partial M$.
\end{Theorem}

\section{Proofs of Theorems~\ref{T1global2} and~\ref{T1global}}
\label{Sproofs}

{\sc Proof of Theorem~\ref{T1global2}:} Let $Q$ be the polynomial
of Theorem~\ref{TnoKIDs}, set
$$\hat \mcV_p=\{ \mbox {vacuum initial data such that
$Q[K,g](p)\ne 0$} \}\;,$$ then $\hat \mcV_p$ is open and contained
in $\mcV_p$. To show density,
 let
$M_i\subset M $ be a sequence of relatively compact domains with
smooth boundary such that $M=\cup_i M_i$. The argument of the
proof of Lemma~\ref{Lrest} shows that $\mcK(M_i)=\{0\}$ for $i$
large enough. Point 1 follows then from Theorem~\ref{Tmain} with
$M$ there equal to $\overline M_i$. The time-symmetric case is
obtained similarly by Theorem~\ref{TmainRiem}. Point 2 is
established by repeating the argument of the proof of
Theorem~\ref{Tsc}. \phantom{xxxxxxxxxxxxxx}\qed

{\noindent \sc Proof of Theorem~\ref{T1global}:} Openness follows
from Proposition~\ref{Pclosed}, it remains to establish density.
We start by showing that for spatially compact CMC initial data
KIDs are ``purely spacelike". Somewhat more generally, one has:

\begin{Proposition}\label{P3}
 Consider a vacuum initial data set $(M,g,K)$
with constant $\tau:=\tr_g K=0$, suppose that\eq{taulamc} holds,
and assume that $(M,g)$ is geodesically complete (perhaps as a
manifold with boundary). Let  $(N,Y)$ be a KID on $M$ satisfying
\bel{ast}\lim_{r\to\infty}\sup_{q\in S_p(r)}|N(q)|=0\;,\ee for some $p\in M$, where
$S_p(r)$ is the boundary of the geodesic ball of radius $r$
centred at $p$.
 If $N\not \equiv 0$, then $K$ is pure trace,
$M$ is compact and $(M,g)$ is Einstein.
\end{Proposition}

\begin{Remark}\label{RP3}
A KID satisfying \eq{ast} will be called \emph{asymptotically
tangential}; a KID with $N\equiv 0$ will be called tangential. In
the compact boundaryless case we have $S_p(r)=\emptyset$ for $r$
large enough, so all KIDs are asymptotically tangential.
\end{Remark}

\proof  We note that if $M$ has a boundary, then $S_p(r)\cap
\partial M :=\partial B_p(r)\cap \partial M \ne \emptyset$ for $r$ large,
so that \eq{ast} implies that $N$ vanishes on $\partial M$. The
KID equations imply
$$\Delta N = \left(|K|^2-\frac {2\Lambda}{(n-1)}\right) N=
\left(|\tilde K|^2+\frac{(\tr_g K)^2}n-\frac
{2\Lambda}{(n-1)}\right) N\;,$$ where $\tilde K$ is the trace-free
part of $K$. \Eq{taulamc} and the maximum principle show that
either $\tilde K\equiv 0$, with \eq{taulamc} being an equality,
and $N=\const$, or $N\equiv 0$. In the former case the KID
equations further imply Ricci flatness of $g$. The case $N\not
\equiv 0$ is compatible with \eq{ast} only if $S_p(r)=\emptyset$
for $r$ sufficiently large, which is equivalent to compactness of
$M$. \qed

We note the following straightforward consequence of
Theorem~\ref{TnoCKV} and Proposition~\ref{Pclosed}:

\begin{Proposition}
\label{P2} Consider the collection of Riemannian metrics on a
three dimensional manifold  with a $C^k$ (weighted, with arbitrary
weights, in the non-compact case) topology, $k\ge 5$. The set of
such Riemannian metrics which have no globally defined conformal
Killing vectors is open and dense.
\end{Proposition}

\proof Choose any relatively compact $\myOmega\subset M$, then by
Theorem~\ref{TnoCKV} and Proposition~\ref{Pclosed} there exists an
open and dense set of metrics which have no conformal Killing
vector fields on $\myOmega$, then those metrics do not have
globally defined conformal Killing vector fields either. \qed

Our next result uses spaces $C^{k,\alpha}_{\varphi,\psi}$ defined
in Appendix~\ref{Atopo}. The weights $\varphi$ and $\psi$ in our
next result have to be chosen in a way compatible with the
conformal method in the asymptotically flat
regions~\cite{christodoulou:murchadha}, similarly in the
asymptotically hyperbolic regions~\cite{AndChDiss}, while
$\varphi=\psi=1$ in the compact case. The differentiabilities here
are different, as compared to Theorem~\ref{T1global}, because
under the CMC restriction the conformal method can be used:

\begin{Corollary}\label{Co1} There exists $k_1(n)$, with $k_1(3)=5$,
such that for  $k\ge k_1(n)$ and $\alpha\in (0,1)$ the following
holds: There exists a $C^{k,\alpha}_{\varphi,\psi}\times
C^{k-1,\alpha}_{\varphi,\psi}$-open and dense collection of vacuum
CMC initial data sets $(M,g,K)$ which are either
\begin{enumerate}
\item asymptotically flat with compact interior (then $\Lambda=0$), or
\item asymptotically hyperbolic as in~\cite{AndChDiss}, or
\item defined on a compact $M$, 
with
$\Lambda$ satisfying \eq{taulamc},
\end{enumerate}and which do not have any asymptotically
tangential KIDs.
\end{Corollary}

\proof Let $\mcU$ be the set of vacuum initial data $(g,K)\in
C^{k,\alpha}_{\varphi,\psi}\times C^{k-1,\alpha}_{\varphi,\psi}$
such that  $g$ is  Einstein. This class of initial data obviously
forms a closed set with no interior. Let $\mcV_1$ be the
complement of $\mcU$ within the set of all vacuum
$C^{k,\alpha}_{\varphi,\psi}\times C^{k-1,\alpha}_{\varphi,\psi}$
initial data, then $\mcV_1$ is open and dense. Choose any $p\in
M$, and let $\mcV_2$ be the set of initial data in $\mcV_1$ such
that $K$ is not pure trace, and such that the polynomial
$Q[H,DH,D^2H]$ of Theorem~\ref{TnoCKV} does not vanish at $p$,
then $\mcV_2$ is open. Consider any $(g,K)$ which is not in
$\mcV_2$ and which is not in $\mcU$, by Proposition~\ref{P2} for
any $\epsilon>0$ there exists a metric $g'(\epsilon )$ which is in
$\mcV_2$  (and therefore has no conformal Killing vectors) such
that $\|g-g'(\epsilon)\|_{C^5_{\varphi,\psi}}\le \epsilon$.
 Using $K$ as the seed solution for the extrinsic curvature,  the
conformal
method~\cite{AndChDiss,Jimconstraints,christodoulou:murchadha}
allows one to solve for a nearby solution
$(M,g(\epsilon),K(\epsilon))$ of the vacuum constraint equations.
Let $(N,Y)$ be a KID for $g(\epsilon)$. By Proposition~\ref{P3}
the KID $(N,Y)$ is tangential, $N\equiv 0$, which implies that
 $Y$ is a Killing vector field of $g(\epsilon)$.
Now $g(\epsilon)$ is a conformal deformation of $g'(\epsilon)$,
therefore $Y$ is a conformal Killing vector field of
$g'(\epsilon)$, hence $Y=0$. It follows that $\mcV_2$ provides the
desired open and dense set.
\qed

On compact boundaryless manifolds all KIDs are asymptotically
tangential, and Theorem~\ref{T1global} is established in this
case.

Consider, next, the asymptotically flat case, with an $r^{-\beta}$
weighted topology,  $\beta \in (0,n-2)$. Recall that we want to
prove density of metrics without KIDs.  For such $\beta$ the
result can be established as follows: consider the set of
solutions of the constraint equations on $\R^3\setminus B(0,R)$,
which approach $(g,K)$ at $S(0,R)$  exponentially fast as
in~\cite[Theorem~6.6]{ChDelayHilbert}, and which are
$r^{-\beta}$-asymptotically flat. A straightforward generalisation
of~\cite[Corollary~6.3]{ChDelayHilbert} applies to this space of
initial data and shows that this collection forms a manifold. It
follows that each linearised solution of the constraint equations
constructed as at the beginning of the proof of
Theorem~\ref{Tmain} is tangent to a curve of solutions, which
coincide with $(g,K)$ away from the asymptotic region
$\R^3\setminus B(0,R)$. This establishes point 1 of
Theorem~\ref{T1global}. We note that the condition $\tr_gK=0$ is
not necessarily preserved by the perturbation just constructed.
However, it follows from the implicit function theorem, or from
the results of Bartnik~\cite{Bartnik84}, that the deformed initial
data set on $\R^3\setminus B(0,R)$ can be deformed in the
associated space-time to obtain a data set with vanishing mean
extrinsic curvature, proving point 2. Point 3 is established as
above using~\cite{ChDelayHilbert} in the $K\equiv 0$ setting.

In the conformally compactifiable case the argument is identical,
based on~\cite[Theorem~6.7]{ChDelayHilbert}.

In the asymptotically flat case with $\beta=n-2$ some more work is
needed. For simplicity we consider only smooth initial data, but
the construction works also in the finite differentiability case.
The idea is to obtain solutions up to kernel using the techniques
of~\cite{ChDelay,CorvinoSchoenprep}, and to show that one can
correct for the kernel by changing the metric in the asymptotic
region, the argument proceeds as follows. Let $\Gamma(R,2R)$ be a
coordinate annulus, with inner radius $R$ and outer radius $2R$,
contained in the asymptotically flat region, let $x=(K,g)$. Let
$\delta x=(\delta K, \delta g)$ be a solution of the linearised
constraint equations supported in $\Gamma(5R/4,7R/4)$, constructed
as at the beginning of the proof of Theorem~\ref{Tmain}, so that
$x_\epsilon= x+\epsilon \delta x$ has no KIDs on $\Gamma(R,2R)$
for all positive $\epsilon$ small enough. By construction
$x_\epsilon$ fails to solve the constraint equations  by
$O(\epsilon^2)$. We use the terminology of~\cite[Sections~8.1 and
8.2]{ChDelay}. Let $Q_0=(m_0,\vec p_0, \vec c_0, \vec J_0)$ denote
the Poincar\'e charges of $x_0=x$, and for $Q$ in a neighborhood
of $Q_0$ let $y_Q=(K_Q,g_Q)$ be a reference family
 of metrics
obtained on $\R^n\setminus B(R)$  as follows: by scaling,
boosting, and space-translating $(K,g)$ one is led to a family of
initial data sets with mass $m$, ADM-momentum $\vec p$, and centre
of mass $\vec c$ covering a neighborhood of $(m_0,\vec p_0, \vec
c_0)$. Choosing $R$ large enough, a construction
in~\cite{schoen:angmom} can be used to deform each of the
solutions obtained so far to initial data sets with arbitrary
angular momentum in a neighborhood of $\vec J_0$.\footnote{The
point of the current construction is to obtain a ``reference
family", as defined in~\cite{ChDelay}, near the initial data we
started with. An alternative way is to first deform the initial
data to data which are exactly Kerr outside of a compact set with
large radius, and use the Kerr family as the reference family.}
One can now glue $x_\epsilon$ with $y_Q$ using the techniques
described in detail in~\cite{CorvinoSchoenprep,ChDelay} obtaining,
for $\epsilon+|Q-Q_0|$ small enough,  on $\Gamma(R,2R)$ a
``solution up-to-kernel"
$z_{\epsilon,Q}=(K_{\epsilon,Q},g_{\epsilon,Q})$ which smoothly
extends across the inner sphere $B(0,R)$ to $x$, which smoothly
extends across the exterior sphere $B(0,2R)$ to $y_Q$, and which
differs from $x_\epsilon$ by terms which are quadratic in
$\epsilon$ and in $Q-Q_0$. Making $  \epsilon$ and $|Q-Q_0|$
smaller if necessary, the arguments presented in Sections 8.1 and
8.2 of~\cite{ChDelay} show that one can find $Q(\epsilon)$ so that
$z_{\epsilon,Q(\epsilon)}$ solves the constraints, providing the
desired solution without global KIDs. \qed

\appendix

\section{Topologies}\label{Atopo}

In this paper we prove both density and openness results, and
there does not seem to be a topology which captures both features
in an optimal way. The aim of this appendix is to discuss those
issues in some detail.

 As already pointed out in the introduction,
a possible topology for which our results hold is the following:
one chooses some smooth complete Riemannian metric $h$ on $M$,
which is then used to calculate norms of tensors and their
$h$-covariant derivatives; we shall denote this topology by
${\mycal T}^k(h)$. If $M$ is compact, the resulting topology is
$h$-independent, and all our results in the compact case hold with
such topologies, for appropriate $k$'s. However, when $M$ is not
compact, there exist choices of $h$ which will lead to different
topologies; nevertheless, for each such choice
Theorems~\ref{T1global2} holds. Further, all the results,
\emph{except the perturbations that remove global KIDs in an
asymptotically flat or asymptotically hyperbolic region}, remain
true if, \emph{e.g.}, weighted $C^{k,\alpha}_{\phi,\varphi}$
topologies defined with respect to $h$ are used, as defined
in~\cite{ChDelay}, with norm
$$
\begin{array}{l}
\|u\|_{C^{k,\alpha}_{\phi,\varphi}(h)}=\sup_{x\in
M}\sum_{i=0}^k\Big(
\|\varphi \phi^i \nabla^{(i)}u(x)\|_h\\
 \hspace{3cm}+\sup_{0\ne d_h(x,y)\le \phi(x)/2}\varphi(x) \phi^{i+\alpha}(x)\frac{\|
\nabla^{(i)}u(x)-\nabla^{(i)}u(y)\|_h}{d^\alpha_h(x,y)}\Big)
\end{array}\;,$$ with \emph{any} weight functions $\phi$ and
$\varphi$; we shall denote such topologies by ${\mycal
T}^{k,\alpha}_{\phi,\varphi}(h)$. Finally, all openness and
density results established in this paper, including statements
involving the field equations, will hold with any choice of $h$
and weight functions except for  the following restriction: if
$(M,g,K)$ contains an asymptotically flat region, and one wishes
to construct a perturbation that gets rid of a globally defined
KID while preserving the field equations, then $h$ should be
chosen to be, {\emph e.g.}, the Euclidean metric in the
asymptotically flat region, with the weights $\phi=r$,
$\varphi=r^{-\beta}$, for some
 $\beta\in (0,n-2]$. Similarly, in the context of
 Corollary~\ref{Co1} and of point 4 of Theorem~\ref{T1global}, the weights in the asymptotically hyperbolic
 region should be chosen in a way compatible with the
 asymptotic conditions in the conformally compactifiable region as in~\cite{AndChDiss}.

While the above topologies seem  satisfactory for most purposes,
the optimal topology for perturbations that get rid, \emph{e.g.},
of Killing vectors, at a given point $p$, is that of convergence
in the space of $k$-th jets of the metric at $p$, with $k\ge
k_0(n)$, for some $k_0(n)$ as described above, on the space of
metrics which coincide with the starting metric $g$ away from a
compact neighborhood of $p$. However, this space is unnecessarily
small for our openness results, which do not hold in such a weak
topology in any case; see also Remark~\ref{Rtopoclosed}.

\section{``Local extends to global" in the simply connected analytic
setting}\label{Sltg} In this appendix we wish to generalise
Nomizu's theorem~\cite{Nomizu} concerning Killing vectors to
conformal Killing vectors and to KIDs. It should be clear that our
argument applies to a large class of similar overdetermined
systems with analytic coefficients, such as, e.g., those
considered in~\cite{BCEG}. In particular the proof given here
applies to Killing vector fields in arbitrary signature, and seems
to be somewhat simpler than the original one.

\begin{Theorem}
\label{Textend} Let $(M,g)$ be a simply connected analytic
pseudo-Riemannian manifold.
\begin{enumerate}
\item Every locally defined conformal Killing vector extends to a
globally defined one.
\item If, moreover, $K$ is also analytic then every locally defined KID extends to a
globally defined one.
\end{enumerate}
\end{Theorem}

\proof We give the proof for KIDs, the argument for conformal
Killing vector fields is identical. Let $r$, $P_\alpha$ and $L_k$
be as in Section~\ref{Spwc}. We note the following:

\begin{Lemma}
\label{Lextend} Consider a KID $x$ defined on an open set
$\myOmega$, let $\gamma:[0,1]\to M$ be a differentiable path such
that $\gamma:[0,1)\to \myOmega$, with $\gamma(1)\not\in
\gamma([0,1))$. Then there exists a neighborhood $\mcU$ of
$\gamma([0,1])$ and a KID $\hat x$ defined on $\mcU$ such that
$x=\hat x$ on $\gamma([0,1))$.
\end{Lemma}

\proof  \Eq{ksys} shows that each covariant derivative $D_\alpha
r$ of $r$ satisfies along $\gamma$ the linear equation
$$\frac{D \left(D_\alpha r\circ \gamma\right) }{ds}= \dot \gamma^\mu \left(\left(D_\mu D_\alpha
r\right) \circ \gamma \right)= \left( P_{\mu\alpha} r\right)\circ
\gamma\;,$$ with the multi-index $\mu\alpha$ in $P_{\mu\alpha}$
defined in the obvious way. It follows that each
$$F_\alpha(s):=\left(D_\alpha r\circ \gamma\right)(s)$$ extends by continuity
to some values, denoted by $F_\alpha(1)$, such that
$$F_\alpha(1)=P_\alpha (\gamma(1))F(1)\;,$$
where $F(1)=\lim_{s\to 1} r(\gamma(s))$. By continuity the
integrability conditions $L_k=0$ are satisfied by $F(1)$, and
therefore, by the argument given after \eq{ksys}, there exists
$\epsilon>0$ and a solution of the KID equations defined on
$B(\gamma(1),\epsilon)$ for some $\epsilon>0$. We can cover
$\gamma([0,1])$ by a finite number of open balls
$B_i:=B(\gamma(s_i),r_i)$, $i=1,\ldots,N$, such that $s_1=0$,
$s_N=1$, $r_N\le \epsilon$, with the balls pairwise disjoint
except for the neighboring  ones: $B_i\cap B_j=\emptyset$ if
$|i-j|>1$. It should be clear that the solution just constructed
on $B(\gamma(1),\epsilon)$ coincides with that which exists
already on the overlap with $B(\gamma(s_{N-1}),r_{N-1})$. The
desired neighborhood is obtained by setting $\mcU= \cup_i
B(\gamma(s_i),r_i)$. \qed

Returning to the proof of Theorem~\ref{Textend}, let $q$ be any
point in $\myOmega$, let $p\in M$, and let $\gamma:[0,1]\to M$ be
any piecewise differentiable path without self-intersections with
$\gamma(0)=q$, $\gamma(1)=p$. Let $I\subset [0,1]$ be the set of
numbers $s$ such that there exists a neighborhood $\mcU_s$ of
$\gamma|_{[0,s]}$ and a KID $x_s$ defined on $\mcU_s$ such that
$x_s=x$ near $p$. Then $I$ is open by definition, it is closed by
Lemma~\ref{Lextend}, therefore $I=[0,1]$. We have thus shown:

\begin{Lemma}
\label{Lext2} For any piecewise differentiable path
$\gamma:[0,1]\to M$ without self-intersections, with $\gamma(0)\in
\myOmega$, there exist a neighborhood $\mcU$ of $\gamma$ and a KID
$x_\gamma$ defined on $\mcU$, coinciding with $x$ on $\mcU\cap
\myOmega$. \qed
\end{Lemma}

Any $\gamma$ as in Lemma~\ref{Lext2} allows us therefore to extend
$x$ to a neighborhood of $p$. It remains to show that this
extension is $\gamma$--independent. Let thus $\gamma$ and $\hat
\gamma$ be two differentiable paths from $q$ to $p$ without
self-intersections, since $M$ is simply connected there exist a
homotopy of differentiable paths $\gamma_t:[0,1]\to M$, $t\in
[0,1]$, with $\gamma_t(1)=p$, $\gamma_t(0)=q$, $\gamma_0=\gamma$
and $\gamma_1=\hat \gamma$. If any $\gamma_t$ self-intersects at
$s_1$ and $s_2$, with $s_1<s_2$, we replace it by a new path,
still denoted by $\gamma_t$, obtained by staying at
$\gamma_t(s_1)$ for $s\in [s_1,s_2]$; this procedure is repeated
until all self-intersections of $\gamma_t$ have been
eliminated.
 Let $r(t)$ denote the value of $r$ at $p$
obtained from Lemma~\ref{Lext2} by following $\gamma_t$, then $r$
is a continuous function of $t$. The set of $t$'s for which
$r(t)=r(0)$ is closed by continuity of $r$, it is open by
Lemma~\ref{Lext2}, hence $r(0)=r(1)$, which establishes
Theorem~\ref{Textend}. \qed

\bigskip

\noindent {\sc Acknowledgements:} We are grateful to L.~Andersson,
A.~Cap, E.~Delay, A.~Fischer, J.~Isenberg, J.~Lewandowski,
M.~McCallum and D.~Pollack for comments, discussions and
suggestions.

\bibliographystyle{amsplain}
\bibliography{
../references/newbiblio,%
../references/reffile,%
../references/bibl,%
../references/Energy,%
../references/hip_bib,%
../references/netbiblio,../references/addon,../references/newbib}

\def\cprime{$'$}
\providecommand{\bysame}{\leavevmode\hbox to3em{\hrulefill}\thinspace}
\providecommand{\MR}{\relax\ifhmode\unskip\space\fi MR }
\providecommand{\MRhref}[2]{%
  \href{http://www.ams.org/mathscinet-getitem?mr=#1}{#2}
}
\providecommand{\href}[2]{#2}
\begin{thebibliography}{10}

\bibitem{AlinhacMetivier}
S.~Alinhac and G.~M{\'e}tivier, \emph{Propagation de l'analyticit\'e des
  solutions de syst\`emes hyperboliques non-lin\'eaires}, Invent. Math.
  \textbf{75} (1984), 189--204.

\bibitem{AndChDiss}
L.~Andersson and P.T. Chru\'{s}ciel, \emph{On asymptotic behavior of solutions
  of the constraint equations in general relativity with ``hyperboloidal
  boundary conditions''}, Dissert. Math. \textbf{355} (1996), 1--100.

\bibitem{Bartnik84}
R.~Bartnik, \emph{The existence of maximal hypersurfaces in asymptotically flat
  space-times}, Comm. Math. Phys. \textbf{94} (1984), 155--175.

\bibitem{bartnik:variational}
\bysame, \emph{Regularity of variational maximal surfaces}, Acta Math.
  \textbf{161} (1988), 145--181.

\bibitem{Bicak:podolsky}
J.~Bi\v{c}\'ak and J.~Podolsk\'y, \emph{The global structure of
  {Robinson-Trautman} radiative space-times with cosmological constant}, Phys.\
  Rev.\ D \textbf{55} (1996), 1985--1993, gr-qc/9901018.

\bibitem{BEM}
J.-P. Bourguignon, D.G. Ebin, and J.E. Marsden, \emph{Sur le noyau des
  op\'erateurs pseudo-diff\'erentiels \`a symbole surjectif et non injectif},
  C. R. Acad. Sci. Paris S\'er. A \textbf{282} (1976), 867--870.

\bibitem{BCEG}
T.~Branson, A.~\v{C}ap, M.~Eastwood, and R.~Gover, \emph{Prolongations of
  geometric overdetermined systems},  (2004), math.DG/0402100v2.

\bibitem{christodoulou:murchadha}
D.~Christodoulou and N.~{\'O} Murchadha, \emph{The boost problem in general
  relativity}, Comm. Math. Phys. \textbf{80} (1980), 271--300.

\bibitem{SCC}
P.T. Chru\'sciel, \emph{On uniqueness in the large of solutions of {E}instein
  equations (``{S}trong {C}osmic {C}ensorship'')}, Australian National
  University Press, Canberra, 1991.

\bibitem{ChDelay}
P.T. Chru\'{s}ciel and E.~Delay, \emph{On mapping properties of the general
  relativistic constraints operator in weighted function spaces, with
  applications}, M\'em.\ Soc.\ Math.\ de France. \textbf{94} (2003), 1--103,
  gr-qc/0301073v2.

\bibitem{ChDelayHilbert}
\bysame, \emph{Manifold structures for sets of solutions of the general
  relativistic constraint equations}, Jour.\ Geom\ Phys. (2004), in press,
  gr-qc/0309001v2.

\bibitem{CIP}
P.T. Chru\'{s}ciel, J.~Isenberg, and D.~Pollack, \emph{Initial data
  engineering},  (2004), gr-qc/0403066.

\bibitem{CollJMP}
B.~Coll, \emph{On the evolution equations for {K}illing fields}, Jour. Math.\
  Phys. \textbf{18} (1977), 1918--1922.

\bibitem{CorvinoSchoenprep}
J.~Corvino and R.~Schoen, \emph{On the asymptotics for the vacuum {E}instein
  constraint equations}, gr-qc/0301071, 2003.

\bibitem{Ebin}
D.G. Ebin, \emph{The manifold of {R}iemannian metrics}, Global Analysis, Proc.
  Sympos. Pure Math., vol.~15, 1970, pp.~11--40.

\bibitem{Hormander}
L.~{H\"ormander}, \emph{The boundary problems of physical geodesy}, Arch. Rat.
  Mech. Analysis \textbf{62} (1976), 1--52.

\bibitem{Jimconstraints}
J.\ Isenberg, \emph{Constant mean curvature solutions of the {Einstein}
  constraint equations on closed manifolds}, Class. Quantum Grav. \textbf{12}
  (1995), 2249--2274.

\bibitem{Lohkamp:Ricci}
J.~Lohkamp, \emph{Metrics of negative {R}icci curvature}, Ann. of Math. (2)
  \textbf{140} (1994), 655--683.

\bibitem{MR54:4541}
V.~Moncrief, \emph{Space-time symmetries and linearization stability of the
  {E}instein equations. {II}}, Jour.\ Math.\ Phys. \textbf{17} (1976),
  1893--1902.

\bibitem{Nomizu}
K.~Nomizu, \emph{On local and global existence of {K}illing vector fields},
  Ann. Math. \textbf{72} (1960), 105--120.

\bibitem{Pirani}
F.A.E. Pirani, \emph{Introduction to gravitational radiation theory}, Lectures
  on general relativity, Brandeis, vol.~1, Prentice Hall, Englewood Cliffs, New
  Jersey, 1965.

\bibitem{schoen:angmom}
R.~Schoen, \emph{in preparation},  (2003).

\bibitem{Thomas}
T.Y. Thomas, \emph{The differential invariants of generalized spaces},
  Cambridge University Press, 1934.

\end{thebibliography}

\end{document}